\documentclass{article}

\usepackage{PRIMEarxiv}

\usepackage[utf8]{inputenc} 
\usepackage[T1]{fontenc}    

\usepackage{hyperref}       
\usepackage{xcolor}
\hypersetup{
    colorlinks,
    linkcolor={red!50!black},
    citecolor={blue!50!black},
    urlcolor={blue!80!black}
}

\usepackage{url}            
\usepackage{booktabs}       
\usepackage{amsfonts}       
\usepackage{nicefrac}       
\usepackage{microtype}      
\usepackage{lipsum}
\usepackage{fancyhdr}       
\usepackage{graphicx}       
\graphicspath{{Fig/}}     

\usepackage{natbib}
\usepackage{amsmath}
\usepackage{amssymb}
\usepackage{siunitx}
\DeclareMathOperator*{\argmin}{argmin}

\usepackage{subfig}

\usepackage{siunitx}

\usepackage{appendix}

\usepackage{authblk}

\title{A reciprocity-aware, low-rank regularization for multidimensional deconvolution
}

\author[1,3]{Fuqiang Chen}
\author[1,3]{Matteo Ravasi}
\author[2,3]{David Keyes}
\affil[1]{Earth Science and Engineering Program\protect\\ Physical Science and Engineering Division\protect \\King Abdullah University of Science and Technology (KAUST)}
\affil[2]{Applied Mathematics and Computational Science Program\protect \\
Computer, Electrical and Mathematical Science and Engineering Division\protect \\ King Abdullah University of Science and Technology (KAUST)}
\affil[3]{Extreme Computing Research Center\protect\\King Abdullah University of Science and Technology (KAUST)}
\begin{document}

\maketitle

\begin{abstract}
This paper presents a novel factorization-based, low-rank regularization method for solving multidimensional deconvolution problems in the frequency domain. In this approach, each frequency component of the unknown wavefield is represented as a complex-valued square matrix and approximated using the product of one rectangular matrix and its transpose. The benefit of such a parametrization is two-fold: first, the size of the unknown matrix is greatly reduced compared to that of the original wavefield of interest (and halved compared to conventional factorization-based, low-rank approximations); second, the retrieved wavefield is implicitly guaranteed to comply with the reciprocity principle, as expected from theory. We further show that the proposed objective function can be successfully optimized using the accelerated proximal gradient algorithm and discuss a robust strategy to define the initial guess of the solution. Numerical examples on synthetic and field data demonstrate the effectiveness of the proposed method in compressing the retrieved Green's function while preserving its accuracy.
\end{abstract}

\keywords{Multidimensional deconvolution \and Reciprocity \and Low-rank regularization}

\section*{Introduction}
Multidimensional deconvolution (MDD) is a seismic processing technique commonly used to suppress the effect of an unwanted overburden. Rooted in Green's function representation theory \citep{amundsen2001, Wapenaar2011Gji}, this technique deconvolves the down- and up-going components of a wavefield recorded at a specific datum to estimate the unknown wavefield --- the so-called Green's function or local reflectivity. Over the years, MDD has found numerous applications in seismic data processing, including demultiple \cite[]{Neut2012, RavasiEage2015, boiero2020, Kumaretal2022, Boieroetal2023, Haacke2023}, interferometric redatuming \cite[]{vanderneutetal2011,Wapenaar2011Gji,Vargasetal2021,Shoheietal2011,ravasi2015gji,9785892}, and imaging \cite[]{Filippoetal2014,Wapenaar2014,Ravasi2016GJI,Joost2017Geo,MatteoandIvan2021Geophysics}. 

MDD is known to be an extremely ill-posed inverse problem due to the band-limited and noisy nature of seismic data, as well as the use of spatially limited acquisition geometries. To address this challenge, prior knowledge of the desired Green's function is often incorporated into the inversion process through regularization and/or preconditioning. For instance, \cite{vanderherrmann2012} proposed solving MDD in the curvelet domain by supposing that the sought-after seismic wavefield can be compactly represented in such a domain. \cite{Chenetal2023} suggested representing the wavefield of interest as a stack of rank-deficient complex-valued matrices, introducing a nuclear norm regularization term to favour the retrieval of Green's functions with a low-rank structure. This approach promotes solutions where the singular values of the unknown matrices are sparse (i.e., few non-zero coefficients), effectively reducing the complexity and the number of free parameters in the retrieved wavefields. Alternatively, \cite{Kumaretal2022} proposed to factorize the frequency-domain Green's functions as the product of two rectangular matrices, with the similar goal of imposing low-rank structure onto the solution of the MDD inverse problem. 

Beyond regularization terms informed by prior assumptions on the structure of the wavefield of interest, an alternative method to regularize MDD is to ensure that the estimated Green's function conforms to fundamental principles of physics: as per the reciprocity theorem in wave propagation, the estimated Green's function in MDD should adhere to this principle. \cite{Joost2017Geo}, \cite{luiken2020}, and \cite{Vargasetal2021} have developed a preconditioning technique based on the reciprocity principle, demonstrating its effectiveness in stabilizing MDD, especially when the separated up- and down-going wavefields exhibit strong and coherent noise.

In this paper, we show how the previously mentioned prior information can be naturally embedded together in a novel parametrization of the low-rank factorization of the MDD solution. The upcoming section begins by formulating a novel factorization-based, low-rank regularization method that naturally fulfills the reciprocity property. Following that, we introduce an algorithm for solving regularized MDD using the proposed parametrization. To conclude, we present three examples to demonstrate the benefits of our approach in solving a variety of MDD problems.

\section*{Theory}
The integral form of MDD can be formulated as follows \citep{amundsen2001,Wapenaar2011Gji}
\begin{equation}
U(t,\mathbf{x}_s, \mathbf{x}_r)=\int_{\partial \Omega } D(t, \mathbf{x}_s, \mathbf{x})\ast X(t, \mathbf{x}, \mathbf{x}_r) d\mathbf{x},
\label{eq:continousMDD}
\end{equation}
where $U(t,\mathbf{x}_s, \mathbf{x}_r)$ is the up-going wavefield from a source $\textbf{x}_s$ to a receiver $\mathbf{x}_r$, $D(t,\mathbf{x}_s, \mathbf{x})$ is the down-going wavefield from a source $\textbf{x}_s$ to a line of receivers $\textbf{x}$ observed at the boundary $\partial \Omega$, and $X(t, \mathbf{x}, \mathbf{x}_r)$ is unknown Green's function from a virtual source $\mathbf{x}_r$ to the same line of receiver $\textbf{x}$ of the down-going wavefield.
Here, $\ast$ represents time convolution; however, as the number of time samples in the discretised versions of the $D$ and $U$ wavefields is always large in practical applications, the time convolution in the right-hand side of equation \ref{eq:continousMDD} is usually calculated using fast Fourier transform, i.e.,
\begin{equation}
U(t,\mathbf{x}_s, \mathbf{x}_r)=\mathcal{F}^{-1}\Bigg(\int_{\partial \Omega } \widehat{D}(\omega, \mathbf{x}_s, \mathbf{x}) \widehat{X}(\omega, \mathbf{x}, \mathbf{x}_r) d\mathbf{x}\Bigg),
\label{eq:continousMDD_semiFX}
\end{equation}
where $\mathcal{F}^{-1}$ denotes the inverse Fourier transform, and $\widehat{D}$ and $\widehat{X}$ represent Fourier transformed versions of ${D}$ and ${X}$. 

Assuming that we are interested in reconstructing the wavefield $\mathbf{X}(t, \mathbf{x}, \mathbf{x}_r)$ from a line of virtual sources $\mathbf{x}_r$, which are placed at the same location of the receivers $\mathbf{x}$, the discrete form of equation \ref{eq:continousMDD_semiFX} can be written as
\begin{equation}
\mathbf{U}=\mathbf{F}^{-1}\widehat{\mathbf{D}}\widehat{\mathbf{X}}.
\label{eq:mmm_mdd_semi}
\end{equation}
Consider a seismic survey composed of $n_s$ shots, with each shot being jointly collocated with $n_r$ receivers. In this context, matrices $\mathbf{U}$ and $\widehat{\mathbf{X}}$ have dimensions $n_t n_s \times n_r$ and $n_\omega n_r \times n_r$, respectively. Here, $n_t$ and $n_\omega$ represent the number of samples along the time and frequency axes. In equation \ref{eq:mmm_mdd_semi}, $\widehat{\mathbf{D}}$ is an operator implementing the discretised version of the spatial integral in equation \ref{eq:continousMDD_semiFX}, and $\mathbf{F}^{-1}$ is an operator that computes the inverse fast Fourier transform over the time axis.

Similarly, equation \ref{eq:mmm_mdd_semi} can be written in the frequency domain as
\begin{equation}
\widehat{\mathbf{U}}=\widehat{\mathbf{D}}\widehat{\mathbf{X}},
\label{eq:mmm_mdd_fx}
\end{equation}
where $\widehat{\mathbf{U}}$ is a matrix of size $n_s \times n_r$ representing a single frequency component of the Fourier transformed $\mathbf{U}$, $\widehat{\mathbf{D}}$ is a matrix of size $n_s \times n_r$ representing a single frequency component of the Fourier transformed $\mathbf{D}$, and $\widehat{\mathbf{X}}$ is a matrix of size $n_r \times n_r$ corresponding to a single frequency component of $\widehat{\mathbf{X}}$. Equations \ref{eq:mmm_mdd_semi} and \ref{eq:mmm_mdd_fx} provide two approaches for estimating the sought-after Green's function either by minimizing the difference between $\mathbf{U}$ and $\mathbf{F}^{-1}\widehat{\mathbf{D}}\widehat{\mathbf{X}}$ or between $\widehat{\mathbf{U}}$ and $\widehat{\mathbf{D}}\widehat{\mathbf{X}}$. These two implementations are commonly known as the time- and frequency-domain MDD, respectively. 

Focusing from now on the latter approach, frequency-domain MDD entails solving a linear inverse problem for any given $\omega$: 
\begin{equation}
\min_{\widehat{\mathbf{X}}} (1/2)\big\Vert\widehat{\mathbf{D}}\widehat{\mathbf{X}}-\widehat{\mathbf{U}}\big\Vert_2^2.
\label{eq:mdd_rd_lsq}
\end{equation}
In the following, we will omit the hat symbol for simplicity in notation. Because the forward modeling in equation \ref{eq:mdd_rd_lsq} is represented by a matrix-matrix multiplication process, matrix-based regularization strategies can be used to mitigate the ill-posedness of MDD. A matrix norm, widely used for this purpose, is the nuclear norm. The nuclear norm of a matrix $\mathbf{X}$ is 
$\vert{\mathbf{X}}\vert_{\ast}=\sum \sigma({\mathbf{X}})$, where $\sigma({\mathbf{X}})$ denotes the singular values of $\mathbf{X}$. 
With the addition of a nuclear-norm regularization, equation \ref{eq:mdd_rd_lsq} becomes 
\begin{equation}
\min_{{\mathbf{X}}} (1/2)\big\Vert{\mathbf{D}}{\mathbf{X}}-{\mathbf{U}}\big\Vert_2^2+\lambda\big\vert{\mathbf{X}}\big\vert_\ast,
\label{eq:mdd_rd_lsq_reg}
\end{equation}
where the positive number $\lambda$ controls the trade-off between minimizing the data fitting error and the nuclear norm of $\mathbf{X}$.
Solving equation \ref{eq:mdd_rd_lsq_reg} by means of iterative solvers involves the repeated computation of the singular value decomposition of $\mathbf{X}$ \citep{Recht2010, Cai2010,toh2010accelerated} which may become computationally expensive for large $\mathbf{X}$ matrices.
In this paper, we take another approach to impose such low-rank prior knowledge onto the solution of MDD by factorizing $\mathbf{X}$ into the product of two rectangular matrices.

\subsection*{Low-rank factorization}
As shown in \cite{Recht2010} and \cite{Kumaretal2022}, an alternative low-rank regularization strategy can be achieved through the following factorization: $\mathbf{X}=\mathbf{L}\mathbf{R}$, where $\mathbf{X}\in \mathbb{C}^{n_r\times n_r}$, $\mathbf{L}\in \mathbb{C}^{n_r\times k}$, and $\mathbf{R}\in \mathbb{C}^{k\times n_r}$. When $k<n_r$, we refer to $\mathbf{L}\mathbf{R}$ as the low-rank representation of $\mathbf{X}$. With this representation, calculating $\mathbf{D}\mathbf{L}\mathbf{R}$ requires fewer operations and less memory usage than calculating $\mathbf{D}\mathbf{X}$. When this parametrization is applied to MDD problem, equation \ref{eq:mdd_rd_lsq} becomes 
\begin{equation}
    \argmin\limits_{\mathbf{L},\mathbf{R}}\,(1/2)\|\mathbf{D}\mathbf{L}\mathbf{R}-\mathbf{U}\|_\mathcal{F}^2.
    \label{eq:mdd_xlr_0}
\end{equation}
where $\|\cdot\|_\mathcal{F}$ is the Frobenius norm. It is crucial to emphasize that factorization satisfying the equation \ref{eq:mdd_xlr_0} is not unique \cite[]{Recht2010}. Additionally, it is important to note that the solution obtained from equation \ref{eq:mdd_xlr_0} ensures its rank is precisely $k$, but it may not be minimized. 
Building upon the equivalency established in \cite{Recht2010}:
\begin{equation}
\|\mathbf{X}\|_\ast=\min\limits_{\mathbf{L}\mathbf{R}=\mathbf{X}}\frac{1}{2}(\|\mathbf{L}\|_\mathcal{F}^2+\|\mathbf{R}\|_\mathcal{F}^2),
\end{equation}
we observe that introducing extra regularization terms based on Frobenius norm for $\mathbf{L}$ and $\mathbf{R}$ in equation \ref{eq:mdd_xlr_0} makes it equivalent to MDD with the regularization based on nuclear norm, as shown in equation \ref{eq:mdd_rd_lsq_reg}.
Consequently, equation \ref{eq:mdd_xlr_0} can be reformulated as:
\begin{equation}
    \argmin\limits_{\mathbf{L},\mathbf{R}}\,(1/2)\|\mathbf{D}\mathbf{L}\mathbf{R}-\mathbf{U}\|_\mathcal{F}^2+\lambda(\|\mathbf{L}\|_\mathcal{F}^2+\|\mathbf{R}\|_\mathcal{F}^2).
    \label{eq:mdd_xlr}
\end{equation}

Equation \ref{eq:mdd_xlr} can be viewed as a regularized least-squares problem, where $\displaystyle f_1(\mathbf{L},\mathbf{R})=(1/2)\|\mathbf{D}\mathbf{L}\mathbf{R}-\mathbf{U}\|_\mathcal{F}^2$ denotes the data fitting term. The regularization parameter $\lambda$ controls the balance between minimizing the data misfit and the Frobenius norm of $\mathbf{L}$ and $\mathbf{R}$. Comparing equation \ref{eq:mdd_xlr} with equation \ref{eq:mdd_rd_lsq_reg}, it becomes clear that both equations equivalently search a solution with a minimum rank structure. Notably, the factorization-based low-rank representation in equation \ref{eq:mdd_xlr} is free of singular value decomposition.

To solve equation \ref{eq:mdd_xlr} using gradient-based method, the gradients of $f_1(\mathbf{L}, \mathbf{R})$ with respect to $\mathbf{L}$ and $\mathbf{R}$ are required. Which can be expressed as
\begin{subequations}
\label{eq:dXlr}
\renewcommand{\theequation}{\theparentequation.\arabic{equation}}
\begin{alignat}{2}
 \nabla_{\mathbf{L}} f_1 =&\frac{\partial f_1}{\partial \mathbf{L}}&&=\mathbf{D}^\mathrm{H}(\mathbf{D}\mathbf{L}\mathbf{R}-\mathbf{U})\mathbf{R}^\mathrm{H},\label{eq:dXl} \\
   \nabla_{\mathbf{R}} f_1 =&\frac{\partial f_1}{\partial \mathbf{R}}&&=\mathbf{L}^\mathrm{H}\mathbf{D}^\mathrm{H}(\mathbf{D}\mathbf{L}\mathbf{R}-\mathbf{U}),\label{eq:dXr}
\end{alignat}
\end{subequations}
where $\mathrm{H}$ denotes the conjugate transpose. The details of the derivation of equations \ref{eq:dXl} and \ref{eq:dXr} can be found in Appendix~\ref{gradLandR}. 
\subsection*{Physics-aware low-rank factorization}
According to the reciprocity property of wave propagation, the retrieved wavefield $\mathbf{X}$ should be a symmetric matrix. In fact, $X_{i,j}$ is the response from the a virtual source at location $\mathbf{x}_j$ to a receiver at location $\mathbf{x}_{r,i}$, which must be the same as $X_{j,i}$, the response from a virtual source at location $\mathbf{x}_i$ to a receiver at location $\mathbf{x}_{r,j}$. This property has proven effective in stabilizing MDD, especially when up- and down-going wavefields contain coherent spurious events with significant magnitude \cite[]{Vargasetal2021}. Motivated by the benefit of enforcing such a constraint on the retrieved Green's function, we define here an alternative low-rank factorization, $\mathbf{X}=\mathbf{Q}\mathbf{Q}^\mathrm{T}$. Compared to $\mathbf{X}=\mathbf{L}\mathbf{R}$, this factorization not only explicitly preserves reciprocity in the solution, but also reduces the memory requirements by half since it stores only $\mathbf{Q}$ instead of both $\mathbf{L}$ and $\mathbf{R}$. Finally, we note that our parametrization resembles that of \cite{Burer2003}, although they used it in a different context, namely semidefinite programming for combinatorial optimization.

The least-squares problem associated with the proposed factorization can be formulated as follows: 
\begin{equation}
    \argmin\limits_{\mathbf{Q}}\,(1/2)\|\mathbf{DQ}\mathbf{Q}^\mathrm{T}-\mathbf{U}\|_\mathcal{F}^2+\lambda\|\mathbf{Q}\|_2^\mathcal{F},
    \label{eq:xqqt_ls}
\end{equation}
where $\lambda$ denotes the regularization parameter. Let $f_2(\mathbf{Q})=(1/2)\|\mathbf{DQ}\mathbf{Q}^\mathrm{T}-\mathbf{U}\|_\mathcal{F}^2$ denote the misfit function of data, its gradient with respect to the $\mathbf{Q}$ is given by: 
\begin{equation}
  \nabla f_2 = \frac{\partial f_2}{\partial \mathbf{Q}}=\Big(\mathbf{D}^\mathrm{H}(\mathbf{DQ}\mathbf{Q}^\mathrm{T}-\mathbf{U})+\big(\mathbf{D}^\mathrm{H}(\mathbf{DQ}\mathbf{Q}^\mathrm{T}-\mathbf{U})\big)^\mathrm{T}\Big)\bar{\mathbf{Q}}, 
  \label{eq:dfdq}
\end{equation}
where $\bar{\mathbf{Q}}$ denotes the conjugate of $\mathbf{Q}$. This is a special case of equation \ref{eq:dXlr}. The derivation details can be found in Appendix~\ref{gradLandR}.

\subsection*{Accelerated proximal gradient descent}

The proximal gradient method has been proven to be efficient in solving regularized least-squares problems, particularly when dealing with regularization terms that are separable and have easy to compute proximal operators, such as the $\mathcal{L}_1$ and Frobenius norms. The proximal gradient method conducts a generalized projection step after the ordinary gradient descent update. A key concept in this method is the proximal operator \citep{Parikh}, which computes the solution to an optimization problem defined as follows:
\begin{equation}
    \operatorname{prox}_g(\mathbf{V})=\argmin\limits_\mathbf{X}{\Big(g(\mathbf{X})+(1/2)\|\mathbf{X}-\mathbf{V}\|_\mathcal{F}^2\Big)}.
    \label{eq:proxdef}
\end{equation}

In this work, the Frobenius norm is used as the additional regularization term in equation \ref{eq:xqqt_ls}, represented by $g(\mathbf{Q})=\lambda\|\mathbf{Q}\|_\mathcal{F}^2$. Here, the hyperparameter $\lambda$ controls the strength of the regularization. The proximal operator for this term is defined as: 
\begin{equation}
    \operatorname{prox}_{\lambda\|\cdot\|_\mathcal{F}^2}(\mathbf{V})=\argmin\limits_{\mathbf{Q}}\Big(\lambda\|\mathbf{Q}\|_\mathcal{F}^2+(1/2)\|\mathbf{Q}-\mathbf{V}\|_\mathcal{F}^2\Big).
\end{equation}
which has a closed-form solution:
\begin{equation}
    \operatorname{prox}_{\lambda\|\cdot\|_\mathcal{F}^2}(\mathbf{V})=\frac{\mathbf{V}}{1+2\lambda}.
    \label{eq:proxofFro}
\end{equation}
Equation \ref{eq:proxofFro} reveals that the proximal operator of the Frobenius norm effectively reduces the magnitude of the input matrix $\mathbf{V}$ towards the origin (zero) by a factor proportional to $\lambda$. In other words, it encourages the elements of the matrix $\mathbf{V}$ to be small, and the extent of such damping is controlled by $\lambda$. If $\lambda$ is large, a stronger damping is enforced on the magnitude of $\mathbf{V}$.

The proximal gradient descent method is known for its relatively slow convergence. An improved variant, known as accelerated proximal gradient \citep{FISTA2009}, enhances the convergence rate. Equation \ref{eq:xqqt_ls} can be thereby solved through the following iterative steps:
\begin{subequations}
\label{eq:accproximalIteration}
\renewcommand{\theequation}{\theparentequation.\arabic{equation}}
\begin{align}
\mathbf{P}^{k} &= {\mathbf{Q}}^{k}-\alpha_k(\mathbf{Q}^{k}-\mathbf{Q}^{k-1}).
\label{eq:accproximalIteration2}
\\
\mathbf{Q}^{k+1} &= \mathrm{prox}_{\eta_k g}\big({\mathbf{P}}^{k}-\eta_k \nabla f_2(\mathbf{P}^k)\big),
\label{eq:accproximalIteration1}
\end{align}
\end{subequations}
where $\nabla f_2(\mathbf{P}^k)$ represents the gradient of data misfit $f_2(\mathbf{P}^k)$ evaluated at $\mathbf{P}^k$, $\eta_k$ denotes the step size, which can be determined using the backtracking algorithm, and $\alpha_k$ is a factor to adjusts the current iteration's update direction. \cite{FISTA2009} employ $\alpha_k=(t_k-1)/t_{k+1}$, where $t_{k+1}=(1+\sqrt{1+4t_k^2})/2$ with $t_1=1$. In the case of $\alpha_k=0$, equation \ref{eq:accproximalIteration} reverts to the standard proximal gradient method. 

In order to compare the results of our proposed method with the more commonly used factorization $\mathbf{X}=\mathbf{L}\mathbf{R}$, we require an algorithm to solve equation \ref{eq:mdd_xlr}. The accelerated proximal gradient algorithm is not well-suited for this purpose. A notable distinction between equations \ref{eq:mdd_xlr} and \ref{eq:xqqt_ls} is that the former involves two unknown matrices. We employ an efficient algorithm proposed by \cite{pock2016siam}. The iterations of such an algorithm are outlined as follows:
\begin{subequations}
\label{eq:accproximalIteration_pock}
\renewcommand{\theequation}{\theparentequation.\arabic{equation}}

\begin{align}
\mathbf{Y}^{k} &= \mathbf{L}^{k}-\alpha_k(\mathbf{L}^{k}-\mathbf{L}^{k-1}),\\
\mathbf{L}^{k+1} &= \mathrm{prox}_{\eta_k g}\big(\mathbf{Y}^{k}-\eta_k \nabla_{\kern -0.1em\mathbf{Y}} f_1(\mathbf{Y}^k,\mathbf{Z}^{k-1})\big),
\label{eq:accproximalIteration1_pock}\\
\mathbf{Z}^{k} &= \mathbf{R}^{k}-\beta_k(\mathbf{R}^{k}-\mathbf{R}^{k-1}),
\label{eq:accproximalIteration2_pock} \\
\mathbf{R}^{k+1} &= \mathrm{prox}_{\tau_k g}\big({\mathbf{Z}}^{k}-\tau_k \nabla_{\kern -0.1em\mathbf{Z}}f_1(\mathbf{Y}^{k},\mathbf{Z}^k)\big),
\end{align}
\end{subequations}
where $\nabla_{\kern -0.1em\mathbf{Y}}f_1$ and $\nabla_{\kern -0.1em\mathbf{Z}}f_1$denotes the gradient of the data misfit function $f_1$ with $\mathbf{Y}$ and $\mathbf{Z}$, respectively, $\eta_k$ and $\tau_k$ represent the step length, while $\alpha_k$ and $\beta_k$ are the so-called correction factors \cite[]{pock2016siam}. Equation \ref{eq:accproximalIteration_pock} can be viewed as a generalized version of accelerated proximal gradient, where the unknowns are updated in an alternating style. To reduce the number of hyperparameters when applying algorithm \ref{eq:accproximalIteration_pock} to solve equation \ref{eq:mdd_xlr}, we set $\alpha_k=\beta_k$ following the settings in the reference paper \cite[]{pock2016siam}.

Finally, it is worth noticing that the two low-rank factorizations used in this work render MDD a nonlinear problem: as such, different initial guesses may lead to different estimates due to the possible presence of multiple local minima. When applying the algorithm described in equation \ref{eq:accproximalIteration} to solve equations \ref{eq:xqqt_ls}, the initial guess plays a key role. While initializing the solution with $\mathbf{X}=\mathbf{0}$ is a reasonable choice for multi-frequency MDD, $\mathbf{Q}=\mathbf{0}$ is unsuitable for MDD with factorization-based, low-rank regularization, as illustrated by equation \ref{eq:dfdq}.
One simple approach to warm start the optimization process is to stack an identity matrix of size $k$ to a zero matrix of size $n_r \times k$ for both the real and imaginary parts of $\mathbf{Q}$, i.e., $\mathbf{Q}[:k,:k]$ is the identity matrix, while the remaining entries are filled with zeros.

Another critical aspect in the application of the proposed method is represented by the choice of the following two parameters: the rank $k$ of Green's function and the factor $\lambda$ for the regularization term. This paper primarily focuses on highlighting the ability of the proposed scheme to approximate Green's functions with low-rank representation. Therefore we do not illustrate how values of $\lambda$ impact the results of MDD. Considering that determining an optimal value for $\lambda$ is computationally expensive, we adopt a pragmatic approach to determine an acceptable $\lambda$. With a small value of $k$, we repeat the experiments with $\lambda_{i+1}=0.1\lambda_{i}$ and choose the $\lambda$ that gives the best results. The selected $\lambda$ is then used for experiments with other values of rank $k$. Nevertheless, it is important to note that a non-zero $\lambda$ is needed to ensure that MDD with the proposed factorization-based low-rank regularization converges to a meaningful solution.

\section*{Examples} 
This section provides a series of increasingly complex examples of seismic demultiple and redatuming to illustrate the advantages of incorporating a reciprocal low-rank factorization into the MDD scheme. These advantages are evident in terms of improved accuracy and reduced memory usage.
\subsection*{OBC redatuming}
\begin{figure}
\centering
      \includegraphics[width=0.4\textwidth]{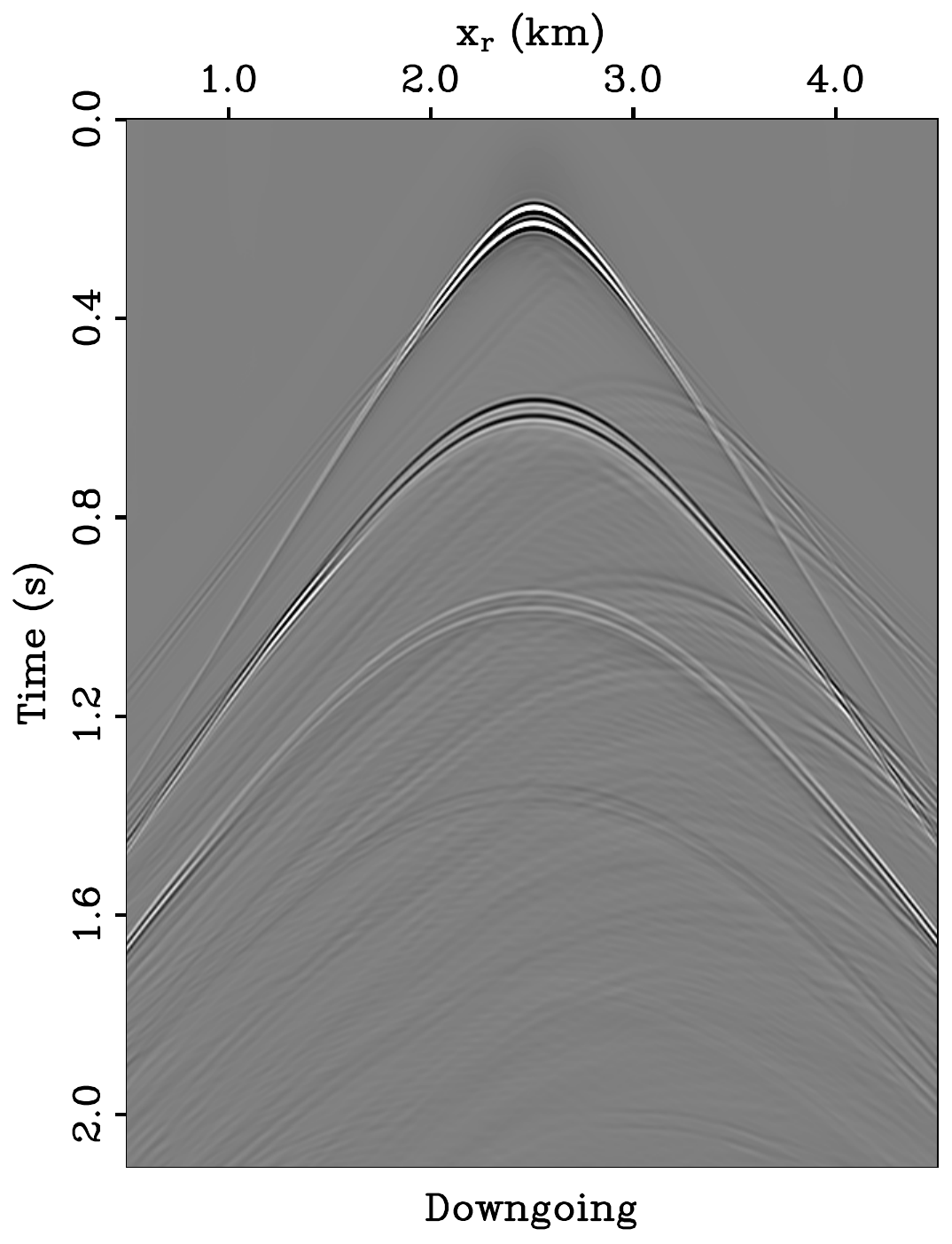}
      \includegraphics[width=0.4\textwidth]{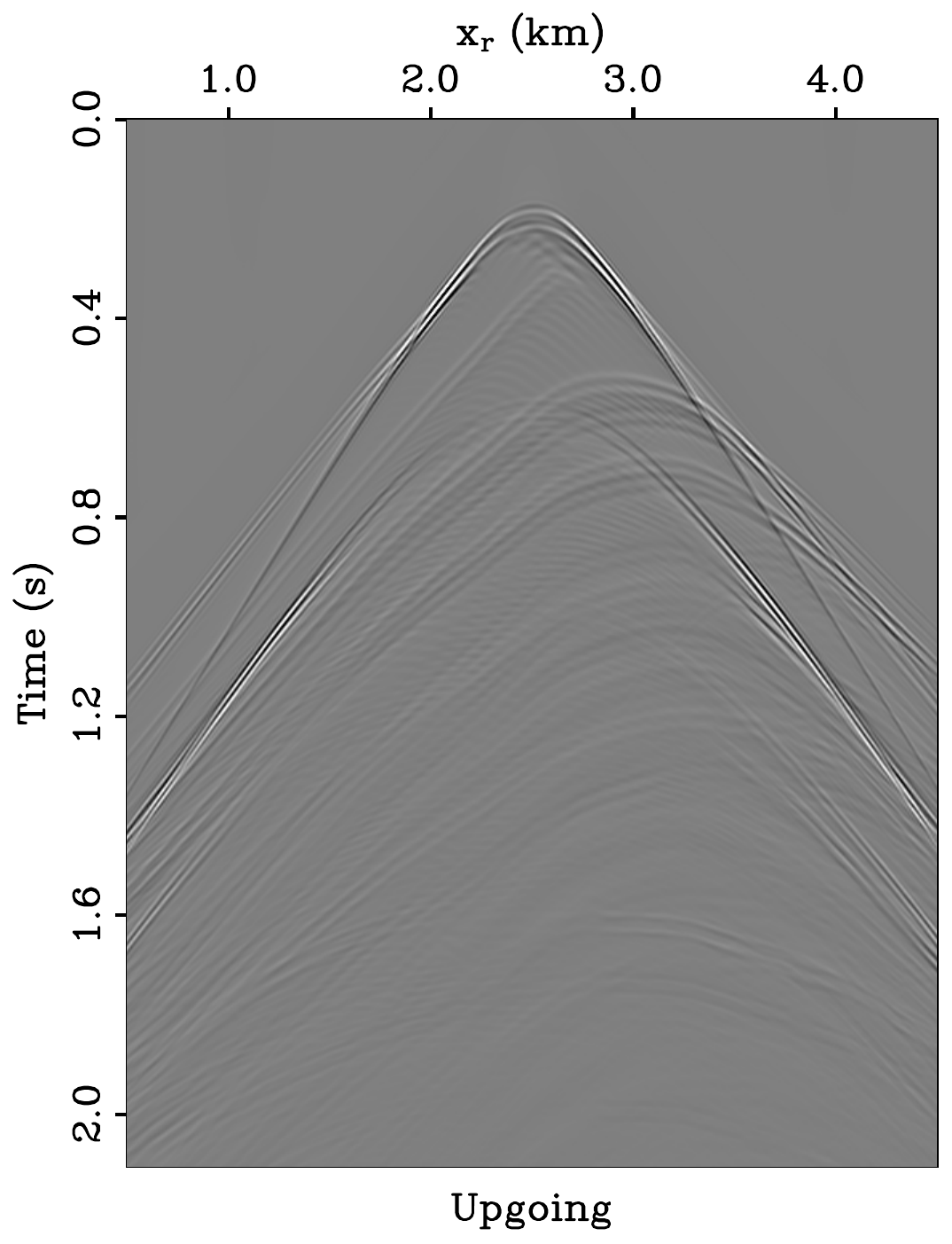}
    \caption{The separated down- and up-going observations at the source location $\mathbf{x}_{\mathrm{s}}=\SI{2.5}{\kilo\meter}$.}
    \label{fig:obc_pdownup}
\end{figure}

We start with a synthetic ocean-bottom cable (OBC) dataset to demonstrate that, when equipped with the suggested low-rank representation, MDD can accurately remove the effect of the free surface from seabed recordings while significantly reducing memory resource demands. In this example, we use a 2D slice of SEG/EAGE Overthrust model \citep{Aminzadeh}, featuring an additional water layer on top. We deploy 401 receivers spaced at an interval of $\SI{10}{\meter}$ on the seafloor, $\SI{292}{\meter}$ away from the free surface, and generate 401 shot gathers with a source spacing of $\SI{10}{\meter}$ and a depth of $\SI{20}{\meter}$. Figure \ref{fig:obc_pdownup} presents an example of the separated down- and up-going observations from a source in the middle of the model used to construct the matrices $\mathbf{D}$ and $\mathbf{U}$. In this illustration, multiple reflections are clearly evident in both the down- and up-going wavefields, resulting from interactions between the free surface and/or the hard seafloor and the subsurface reflectors. Removing these multiples is a crucial step in seismic data processing of OBC data, and MDD offers an effective method for accomplishing this task.

\begin{figure}
\centering
      \includegraphics[width=0.65\textwidth]{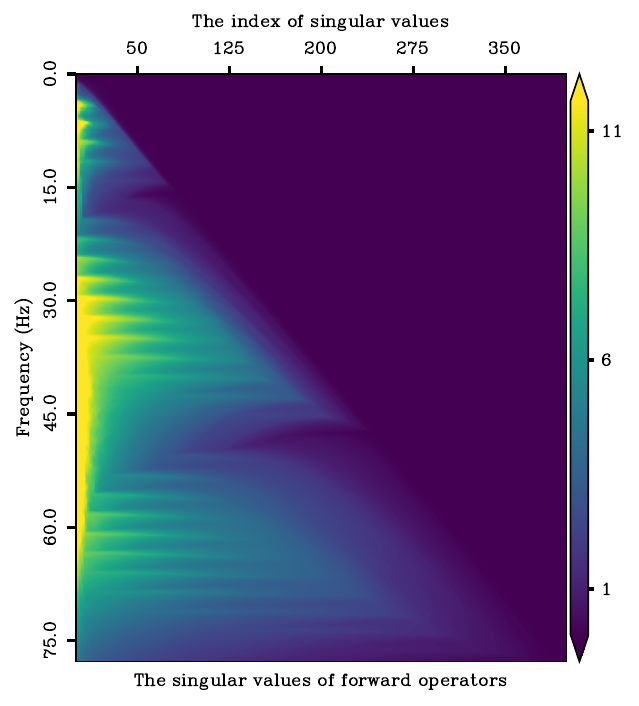}
    \caption{Singular values of forward operator $\mathbf{D}(\omega)$.}
    \label{fig:svalsofA}
\end{figure}

First, to justify the use of a low-rank factorization in MDD, we display the singular values of the forward operator $\mathbf{D}$ in Figure \ref{fig:svalsofA}. This figure reveals that the rank of $\mathbf{D}$ becomes increasingly deficient as the frequency decreases. This trend suggests that the unknown wavefield must exhibit similar properties and, therefore, it can be successfully represented by a low-rank matrix without loss of information. 

\begin{figure}
\centering
    \subfloat[\label{fig:xx}]{
      \includegraphics[width=0.3\textwidth]{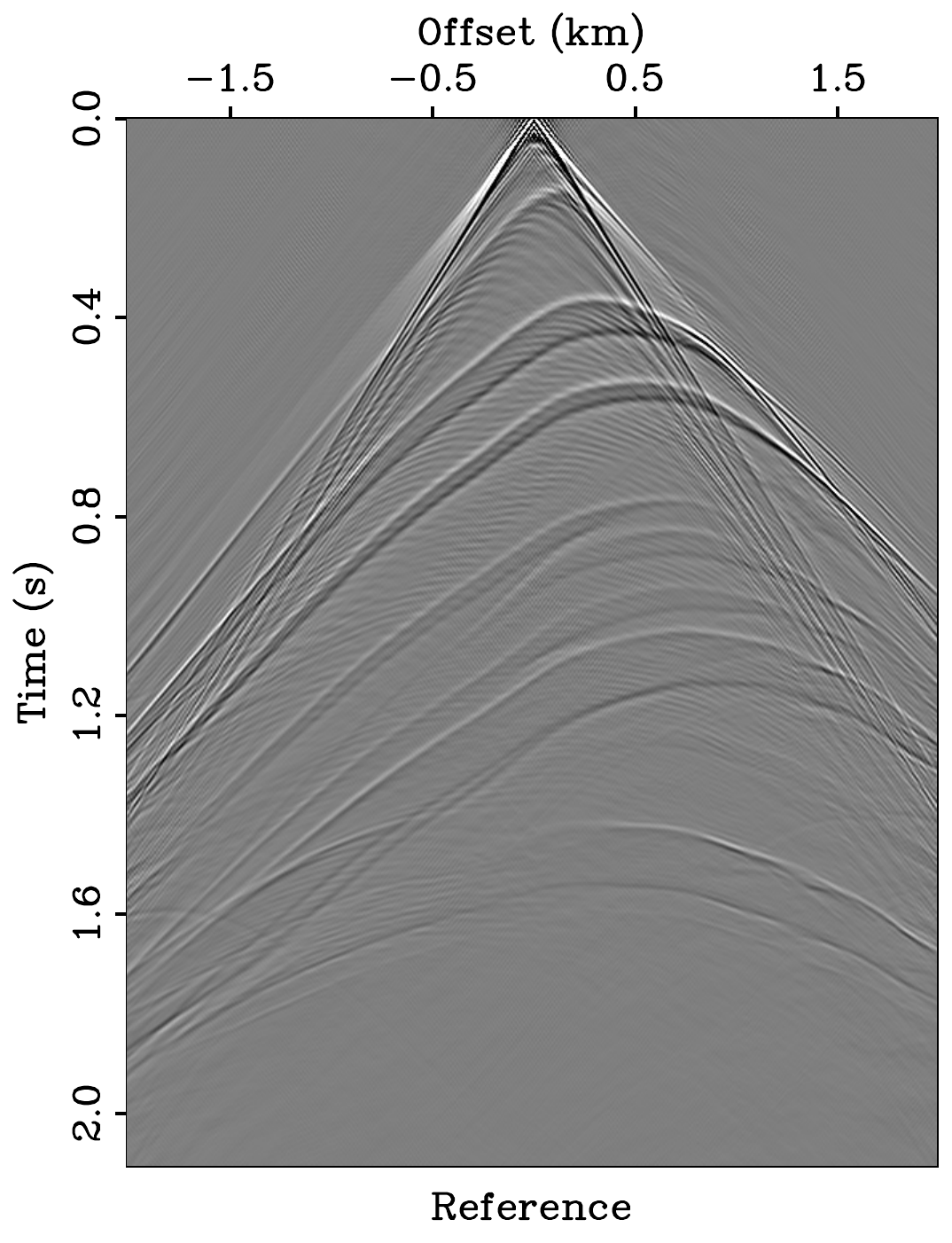}
   }
   \subfloat[\label{fig:yy}]{
      \includegraphics[width=0.3\textwidth]{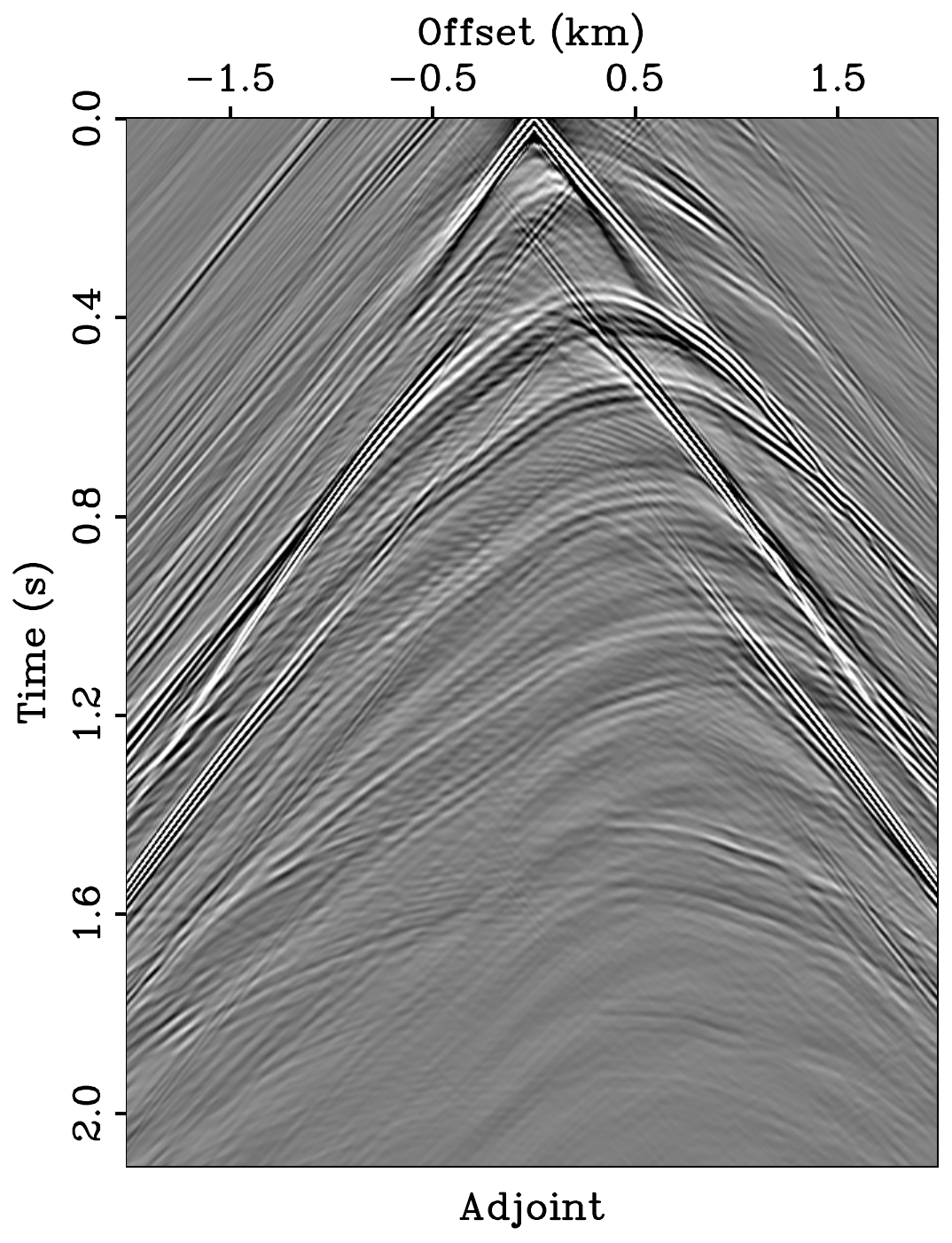}
   }\\
   \subfloat[\label{fig:yy}]{
      \includegraphics[width=0.3\textwidth]{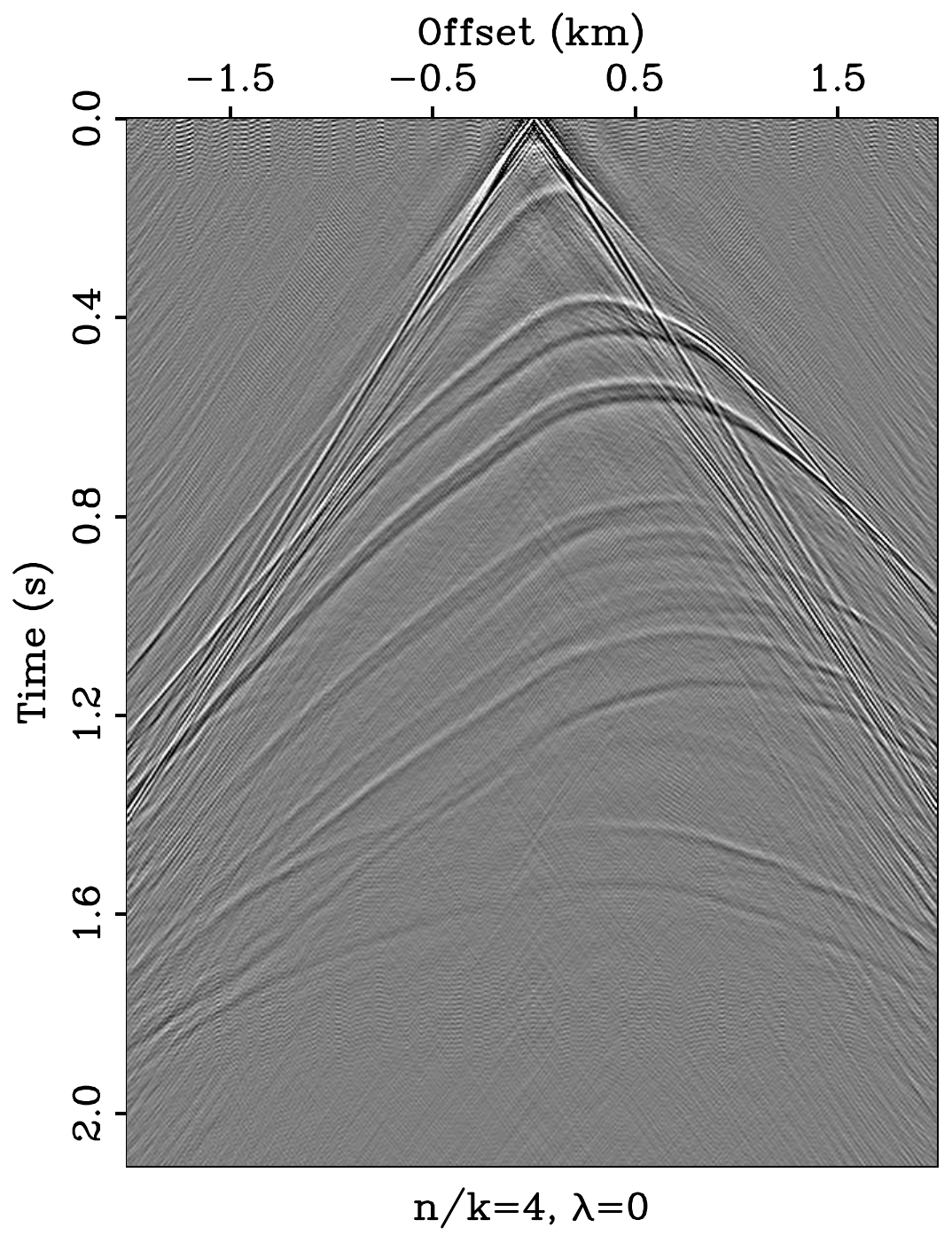}
   }
   \subfloat[\label{fig:yy}]{
      \includegraphics[width=0.3\textwidth]{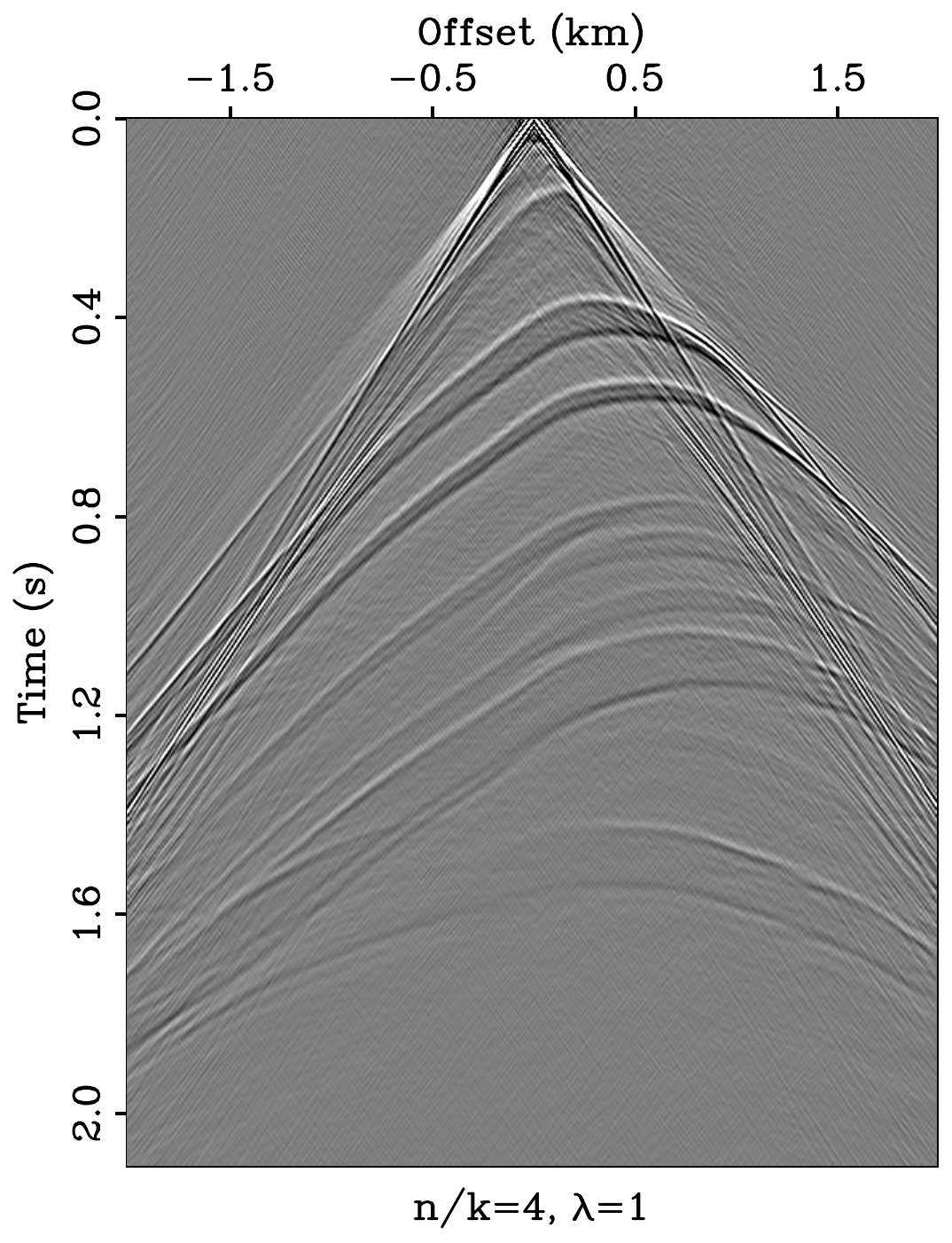}
   }  
    \caption{MDD results from (a) multi-frequency inversion, (b) cross-correlation, (c-d) the proposed method with $k=100$ but different values of $\lambda$.}
    \label{fig:obc_rdtm1}
\end{figure}

Figure \ref{fig:obc_rdtm1}a displays the inversion results obtained through multi-frequency MDD, serving here as the benchmark solution. Note that, since the wavefield separation process does not introduce significant noise in the down- and up-going wavefields, we expect the solution of an unregularized MDD process to be already of reasonably good quality in this example. When compared to the up-going wavefield in Figure \ref{fig:obc_pdownup}, it becomes clear that the Green's function retrieved by MDD does not contain the multiple arrivals associated with the free surface. Figure \ref{fig:obc_rdtm1}b shows the adjoint result, specifically the cross-correlation between down- and up-going wavefields. In contrast to the inversion result shown in Figure \ref{fig:obc_rdtm1}a, it is apparent that the adjoint solution introduces serious artifacts and does not effectively deconvolve the wavelet effect. To demonstrate the importance of choosing an appropriate value for $\lambda$ in improving the quality of our method's performance, Figure \ref{fig:obc_rdtm1}c and \ref{fig:obc_rdtm1}d present the results of our proposed low-rank approximation with $k=100$ but with varying values of $\lambda$ ($\lambda=0$ and $\lambda=1$, respectively). We observe that the result with $\lambda=0$ exhibits appreciable artifacts. Upon comparing this inversion result in Figure \ref{fig:obc_rdtm1}d with the benchmark in Figure \ref{fig:obc_rdtm1}a, we can see that our approach effectively recovers all of the subsurface-born reflection events, removing the scatterings originating from the seafloor. Considering the number of columns of the matrix $\mathbf{X}$ is $n_r=401$, our method reduces the memory requirement for storing the unknown matrix by a factor of $n_r/k=4$.


\subsection*{Target-oriented redatuming}
\begin{figure}
    \centering
    \includegraphics[width=0.8\textwidth]{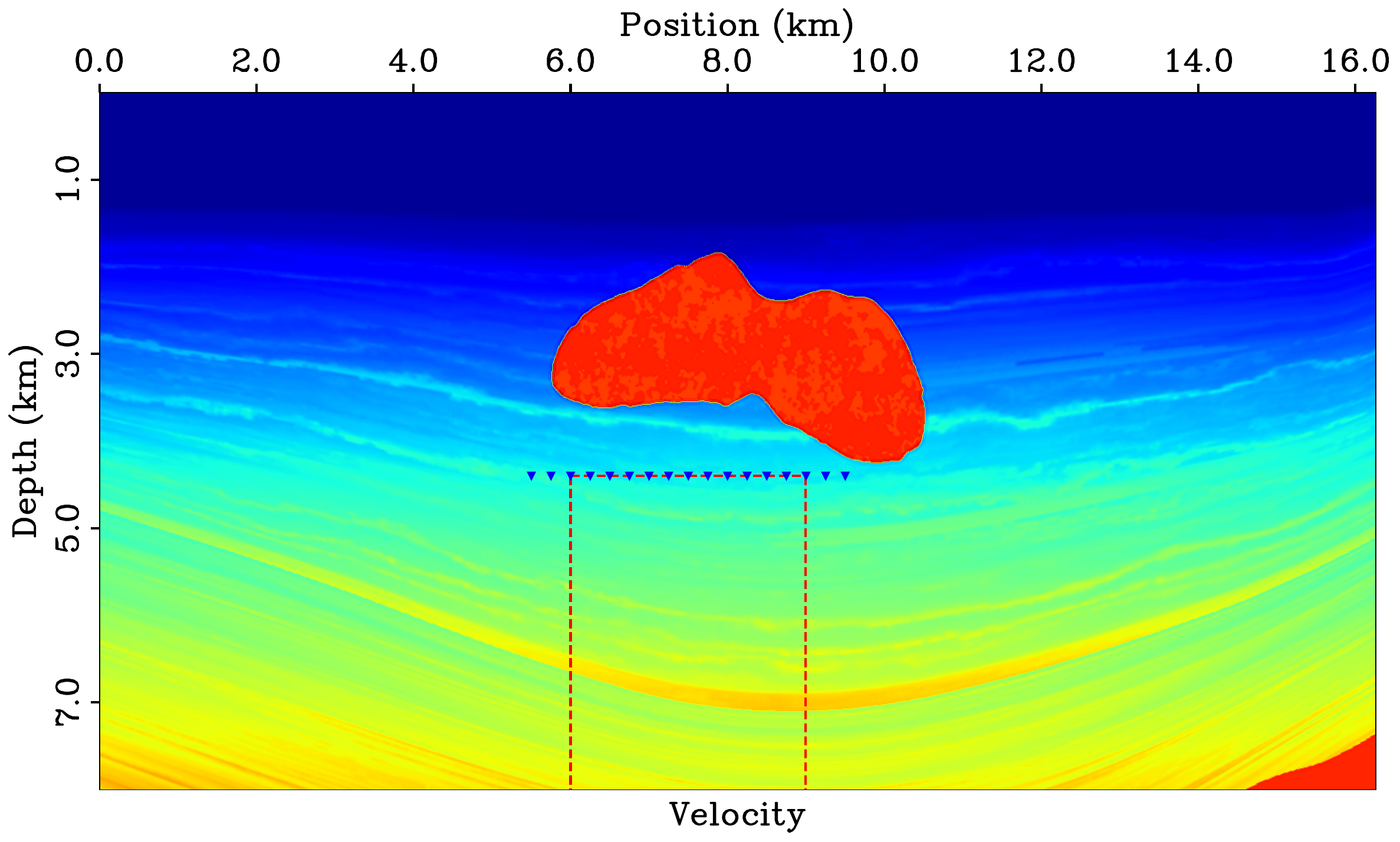}
    \caption{Velocity model to generate data for target-oriented redatuming.}
    \label{fig:subsalt_mdl}
\end{figure}

The second example is built upon a complex velocity model that includes a salt body, as shown in Figure \ref{fig:subsalt_mdl}. The data acquired from surface seismic surveys are known for exhibiting significant scattering, mainly because of the presence of the salt body within this velocity model. These scattering effects make the data analysis and its application for subsurface imaging challenging. Our approach involves the reconstruction of seismic data, where virtual sources and receivers are strategically positioned beneath the salt dome, marked by blue triangles. As a result, the reconstructed data is now free from the interference caused by these scatterings.

In this particular instance, the up- and down-going wavefields are created through the utilization of scattering Rayleigh-Marchenko redatuming \cite[]{vargasetal2021Geophysics}, yielding 201 virtual shot gathers with a source spacing of \SI{40}{\meter}. Each shot gather is equipped with 151 receivers, spaced at \SI{20}{\meter} intervals. Our primary objective in conducting MDD is to recover Green's functions that are specific to the physical properties of the target area, delineated by the dashed red lines in Figure \ref{fig:subsalt_mdl}. Consequently, our seismic observations do not include reflections originating from the free surface or salt dome. It is worth noting that these separated data contain non-trivial spurious events that are unrelated to reflectors in the target area; as such, applying a conventional MDD algorithm without additional constraints would introduce large artifacts in the retrieved solution \citep{Vargasetal2021}.

\begin{figure}
\centering
    \subfloat[\label{fig:subsalt_down_up}]{%
      \includegraphics[width=0.4\textwidth]{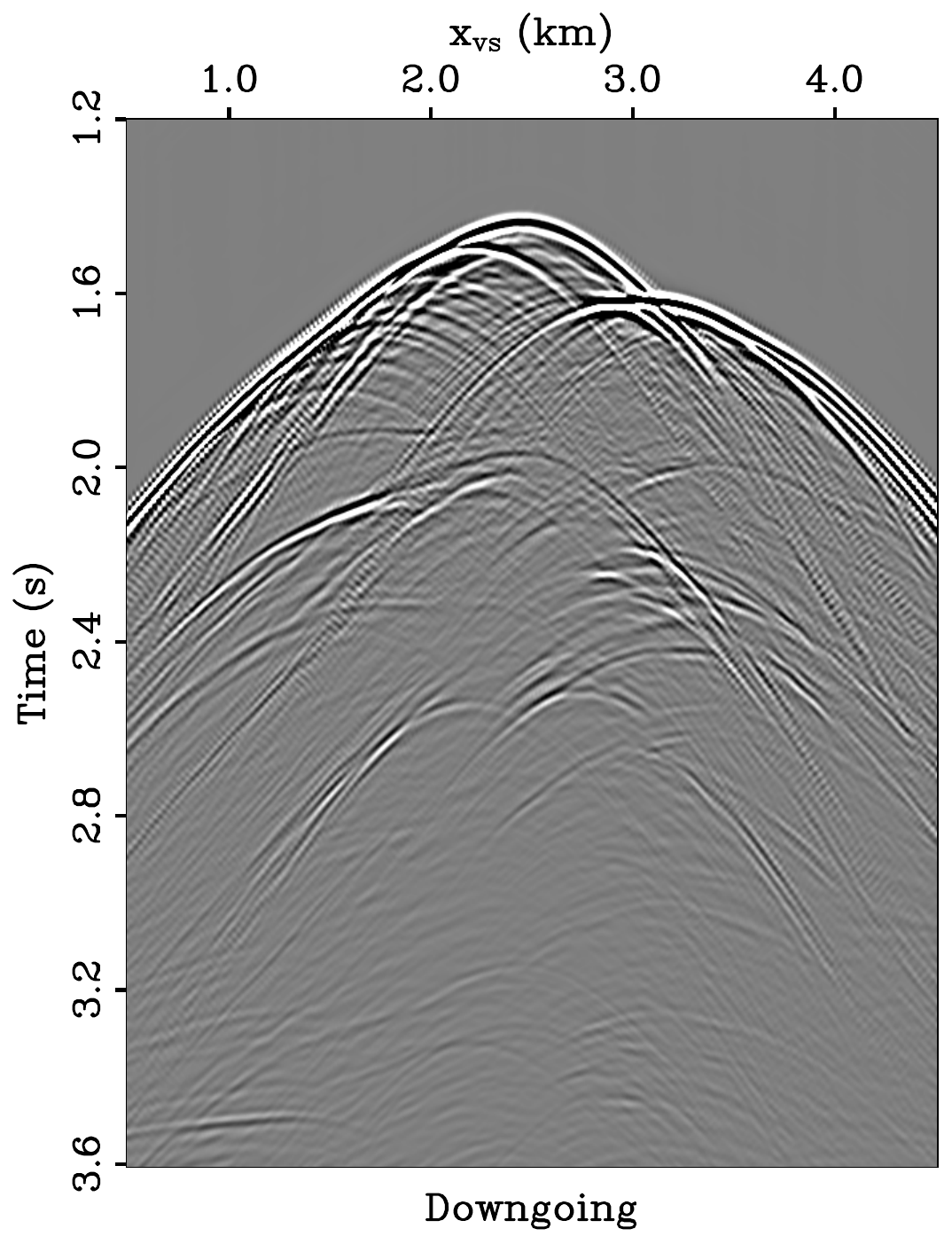}
      \includegraphics[width=0.4\textwidth]{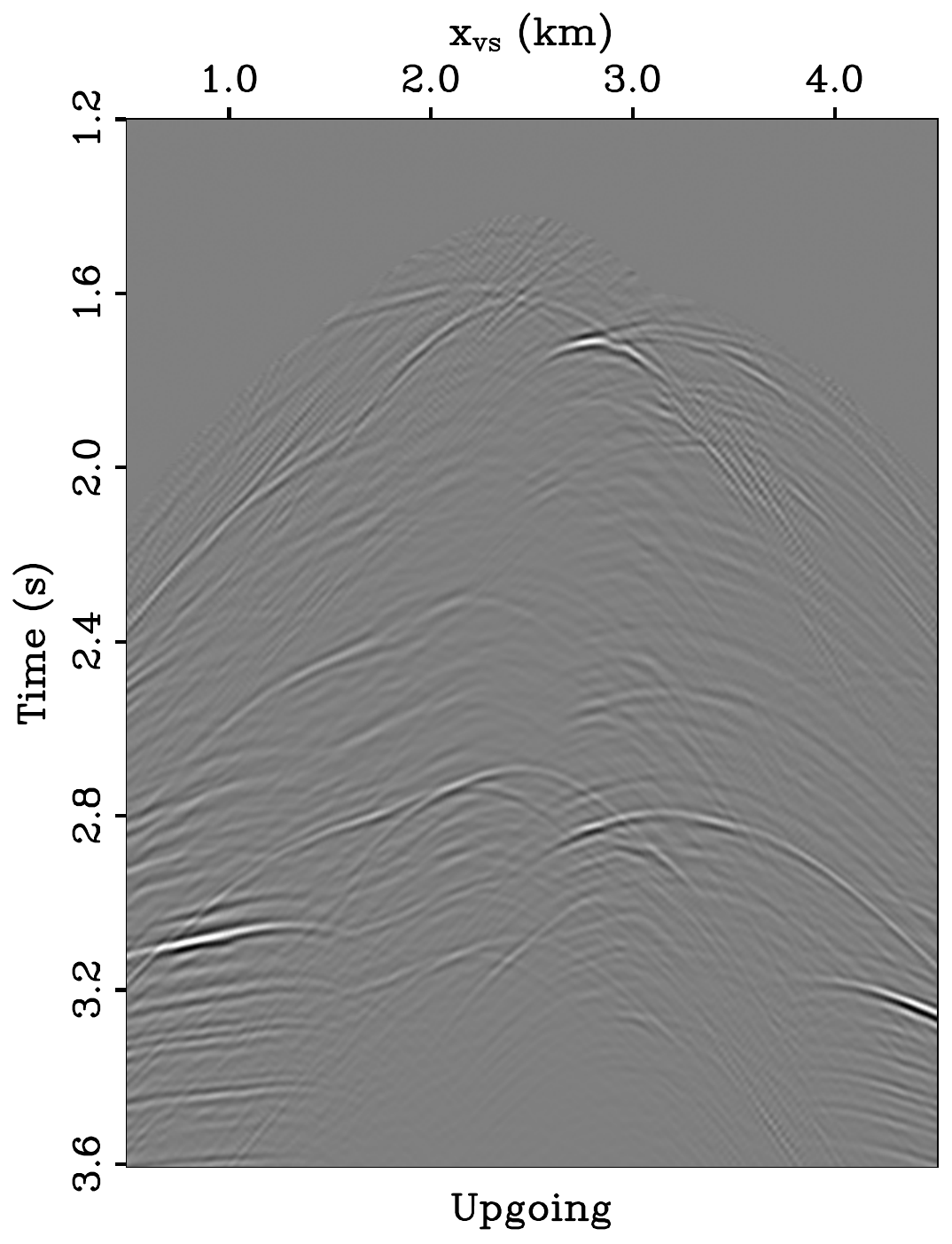}
    }
    \caption{The separated down- and up-going observations at the receiver location $\mathbf{x}_{\mathrm{r}}=\SI{2.5}{\kilo\meter}$.}
    \label{fig:subsalt_k50}
\end{figure}

\begin{figure}
\centering
    \subfloat[\label{fig:subsalt_true}]{%
      \includegraphics[width=0.4\textwidth]{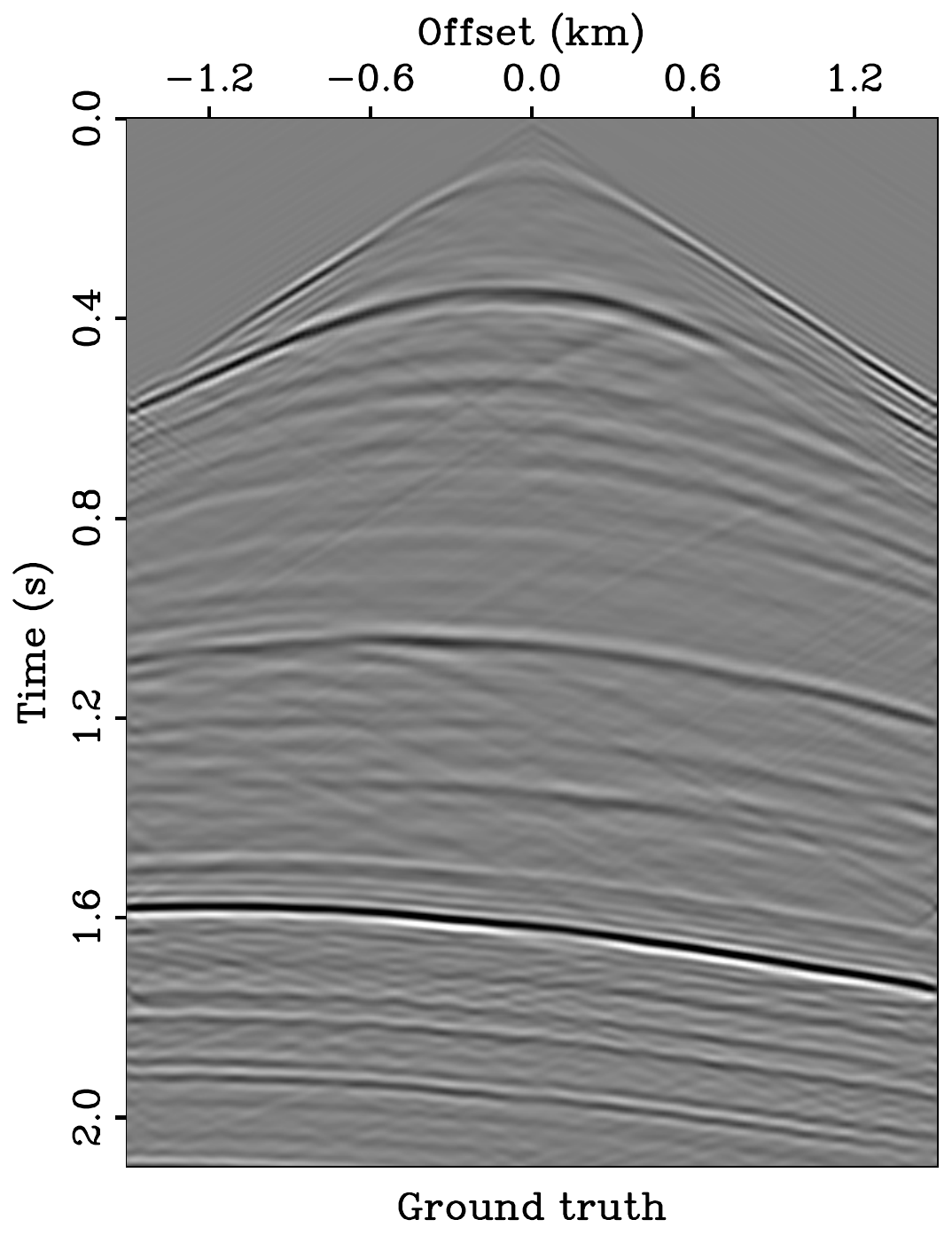}
    }
        \subfloat[\label{fig:subsalt_ref}]{%
      \includegraphics[width=0.4\textwidth]{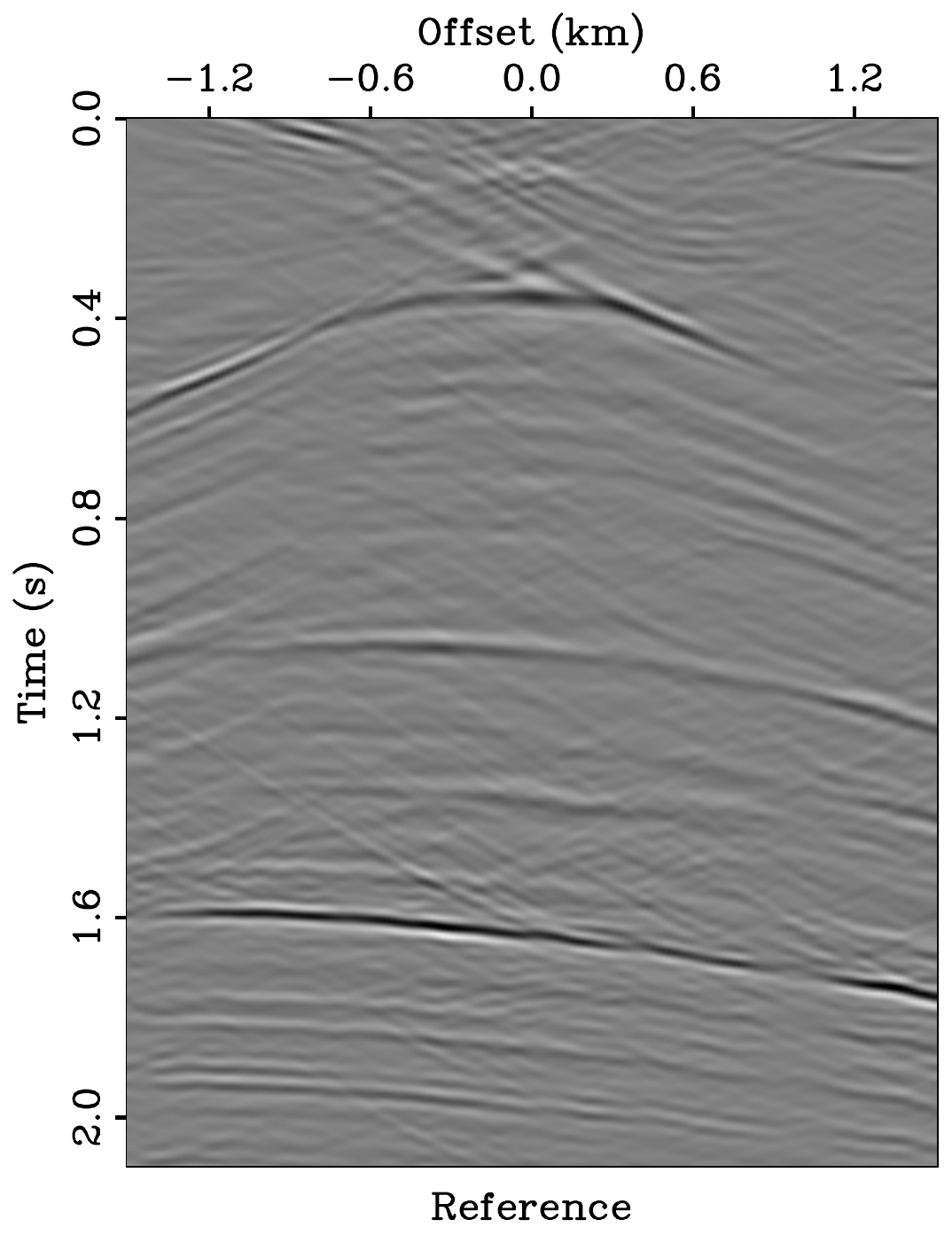}
    }
    \hfill
\centering
    \subfloat[\label{fig:subsalt_xlr}]{%
      \includegraphics[width=0.4\textwidth]{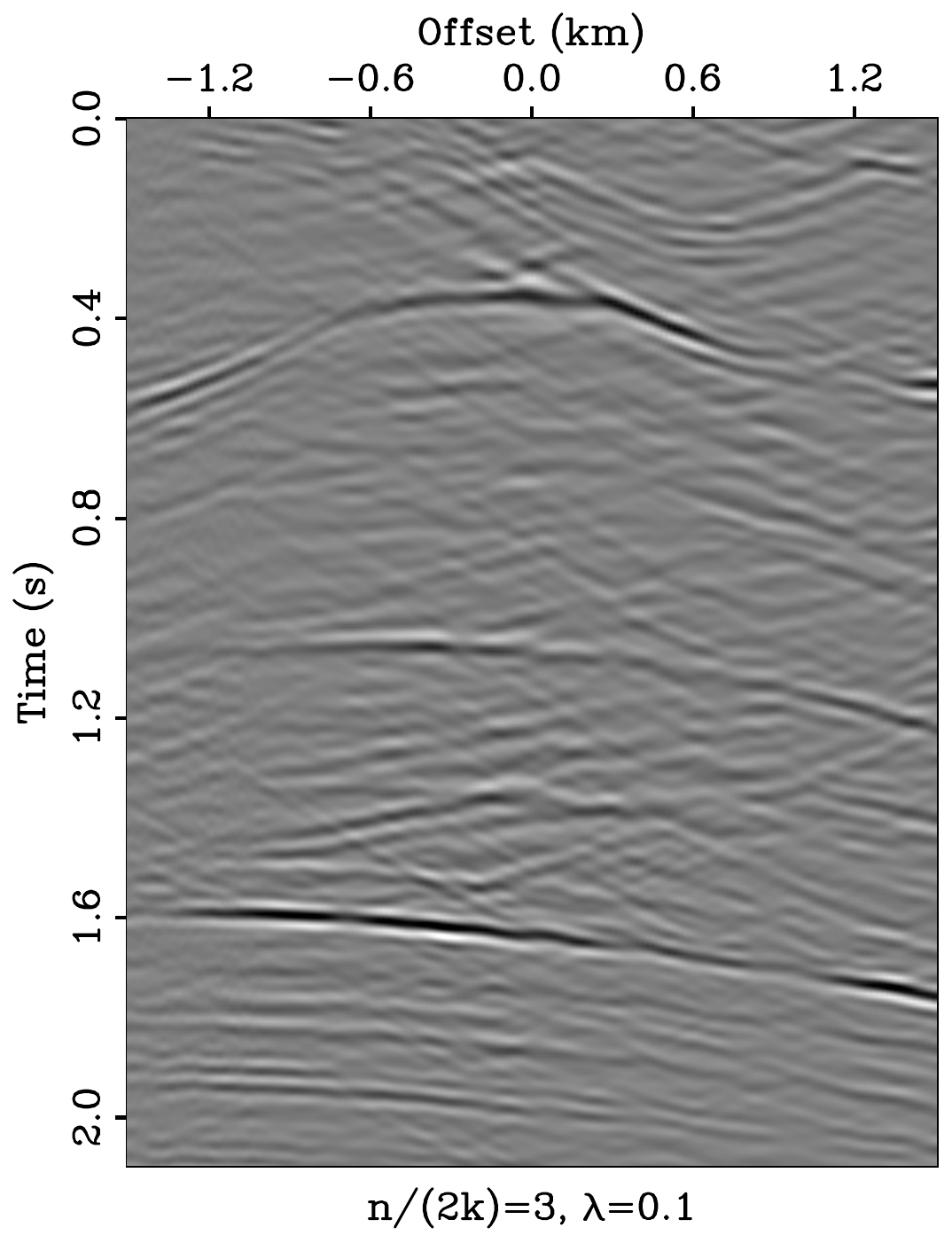}
    }
    \subfloat[\label{fig:subsalt_qqt}]{%
      \includegraphics[width=0.4\textwidth]{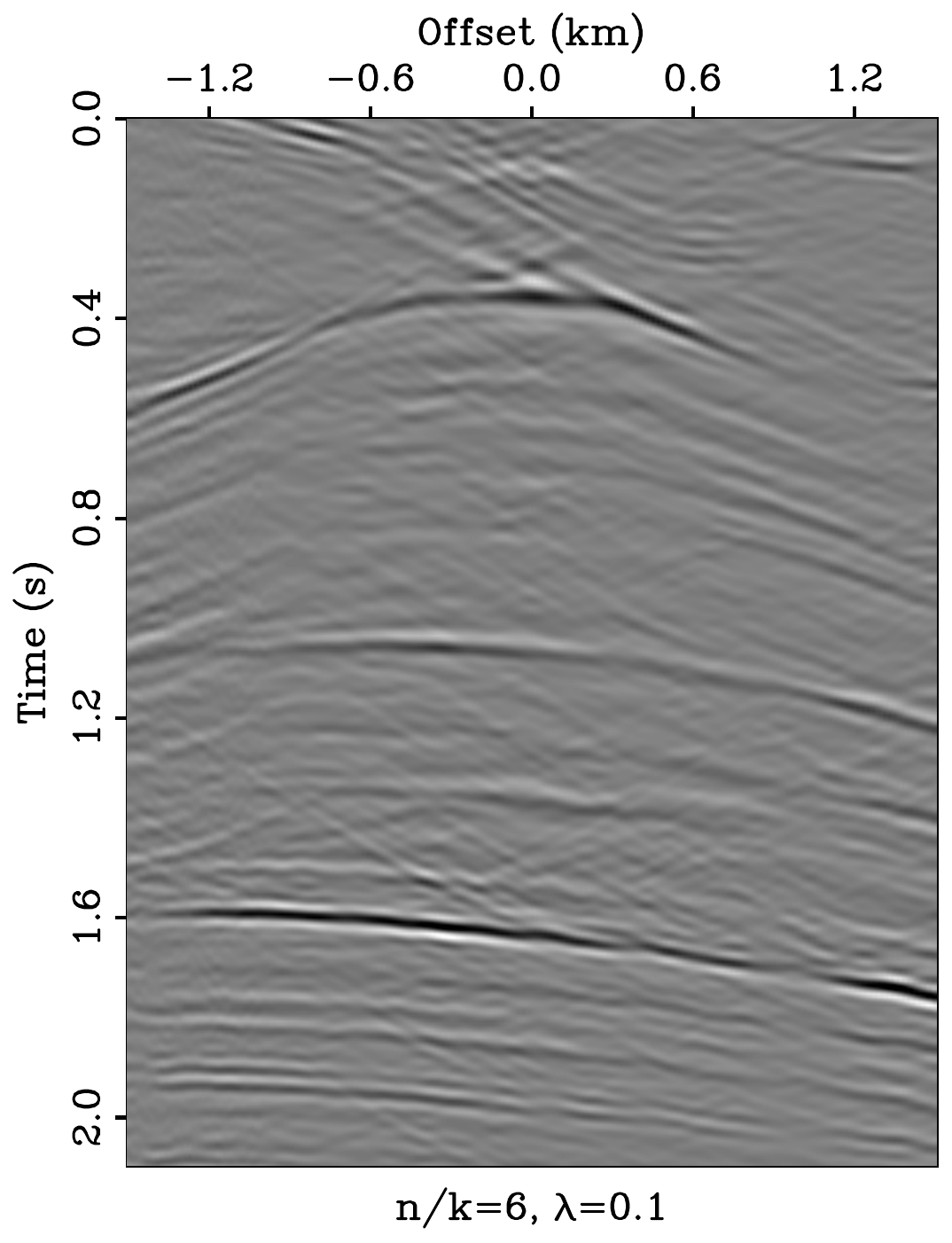}
    }
    \label{fig:subsalt_k50}
    \caption{(a) The true Green's function, (b) the reference from multi-frequency MDD, and the estimated results by MDD regularized by factorization-based low-rank strategy (c) without and (d) with reciprocity constraints.}
\end{figure}

Figure \ref{fig:subsalt_down_up} shows one common receiver gather for separated down- and up-going observations, which are used to construct the matrices $\mathbf{D}$ and $\mathbf{U}$. Notably, scattering signals attributed to the salt body are evident in both the down-going and up-going observations. Figure \ref{fig:subsalt_true} and \ref{fig:subsalt_ref} show one common virtual source gather ($\mathbf{x}_{vs}=\SI{2.5}{\kilo\meter}$) for the true Green's function modelling via finite-difference in the target area and the benchmark results from multi-frequency MDD, respectively. 

To highlight the advantages of the reciprocal low-rank factorization, Figures \ref{fig:subsalt_xlr} and \ref{fig:subsalt_qqt} show the results of MDD using the low-rank factorization $\mathbf{X}=\mathbf{L}\mathbf{R}$ \cite[]{Kumaretal2022} and the proposed one, respectively. In both cases, we choose the same rank value ($k=25$). With this configuration, the proposed low-rank representation and the non-reciprocal approach $\mathbf{X}=\mathbf{L}\mathbf{R}$ will save the memory for storing the Green's function by a factor of $n_r/k=6$ and $2(n_r/k)=3$, respectively. A comparison of our proposed scheme with the non-reciprocal one shows two noteworthy benefits: firstly, our method demonstrates a significant reduction in memory usage, requiring only half the memory compared to the non-reciprocal factorization, while maintaining the same rank value $k$. Secondly, our approach proves to be more robust in obtaining accurate results in this example due to satisfying the reciprocity property. In summary, our proposed method outperforms the non-reciprocal low-rank representation, in terms of enhancing MDD accuracy through low-rank regularization. 

\subsection*{Field data redatuming}
In the final example, we apply the proposed MDD method to the up/down separated wavefields of a 2d line of the Volve field dataset consisting of $119$ shots, each equipped with 180 receivers spaced at regular intervals of \SI{25}{\meter}. Our primary goal here is to generate seismic data that are free from reflections associated with the free surface. More details on the pre-processing steps performed on this dataset can be found in \cite{RavasiEage2015} and \cite{9785892}.

\begin{figure}
\centering
      \includegraphics[width=0.4\textwidth]{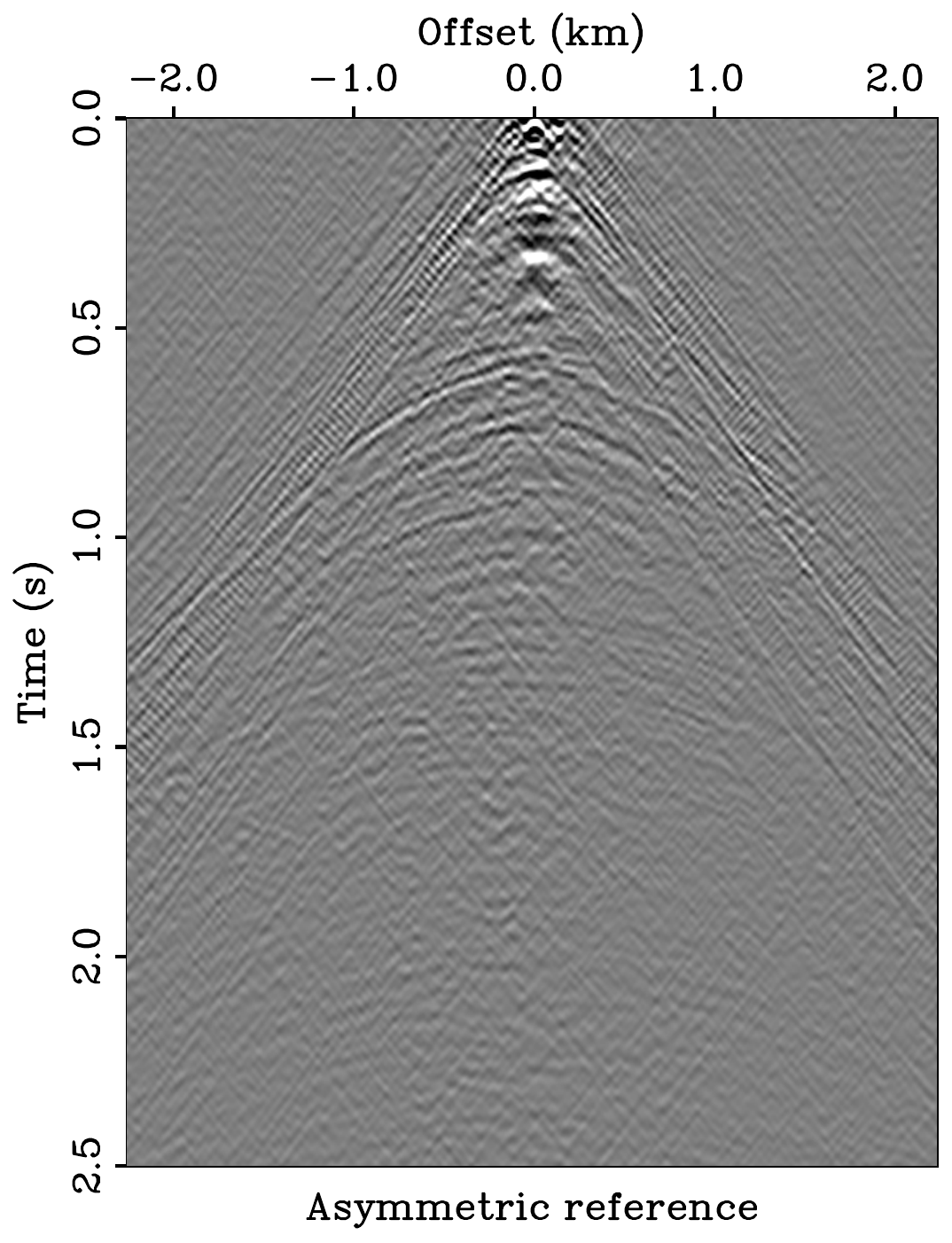}
      \includegraphics[width=0.4\textwidth]{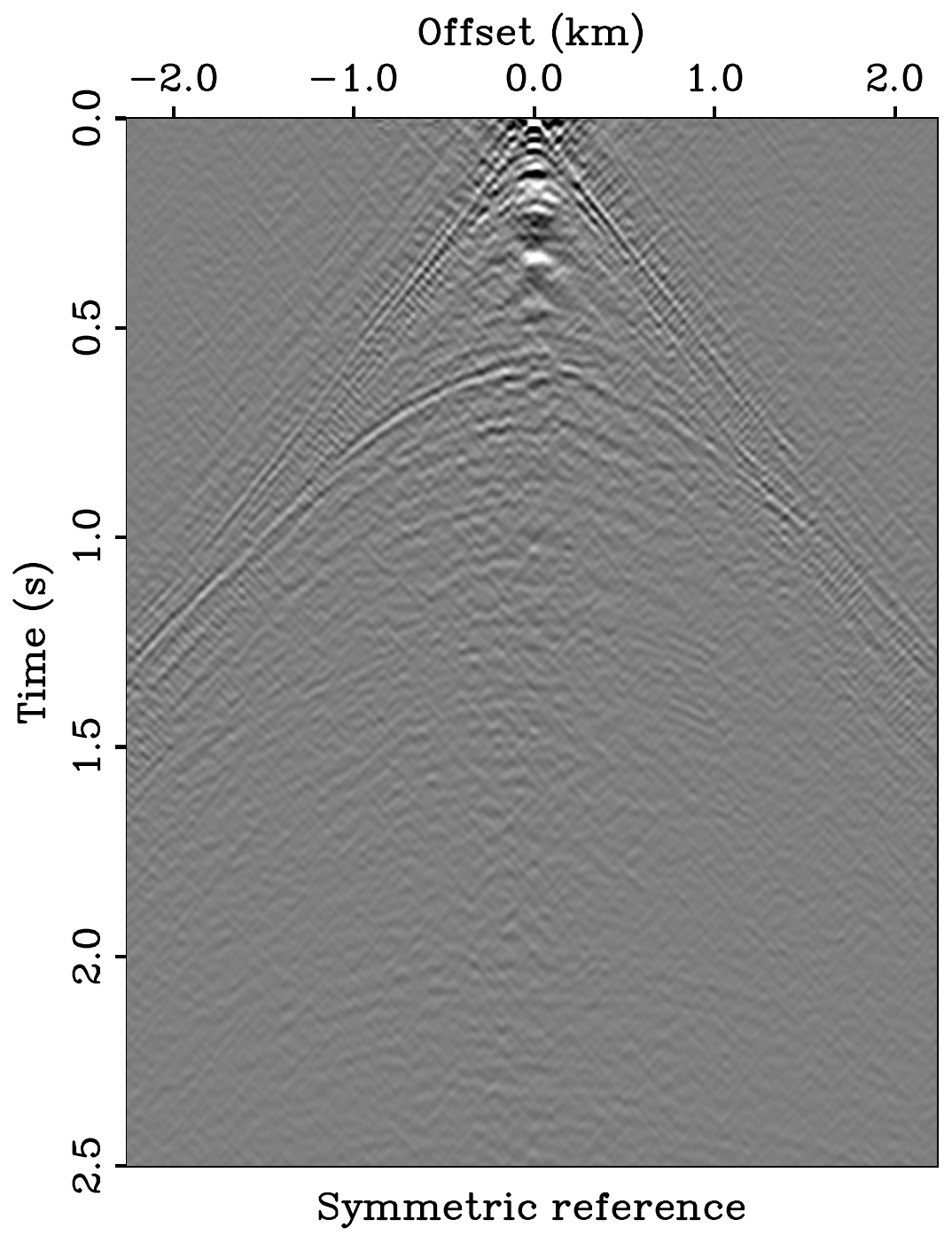}
    \caption{MDD from multi-frequency inversion with and without the reciprocity constraint.}
    \label{fig:Volve_AXB}
\end{figure}

To begin, we create the benchmark results for comparison with the outcomes obtained through our proposed method. To underscore the advantages of incorporating the reciprocity property into MDD, Figure \ref{fig:Volve_AXB} presents time-offset domain results of multi-frequency MDD, without and with reciprocity constraints. 
We refer to the estimated Green's functions without reciprocity as ``asymmetric reference" and those with reciprocity as ``symmetric reference." We can see that results from MDD without reciprocity present noticeably more noise than those where reciprocity is enforced.

\begin{figure}
\centering
      \includegraphics[width=0.4\textwidth]{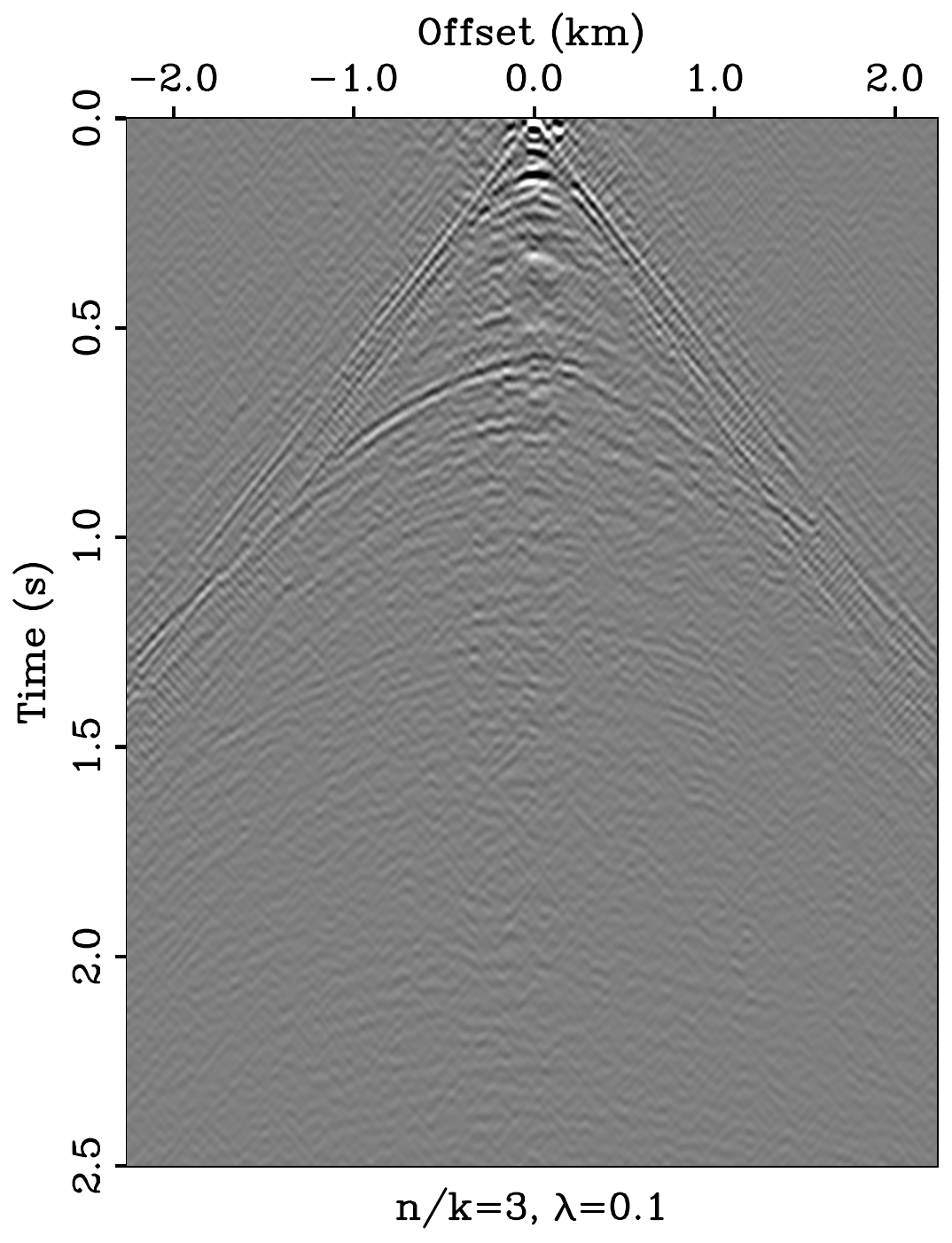}
      \includegraphics[width=0.4\textwidth]{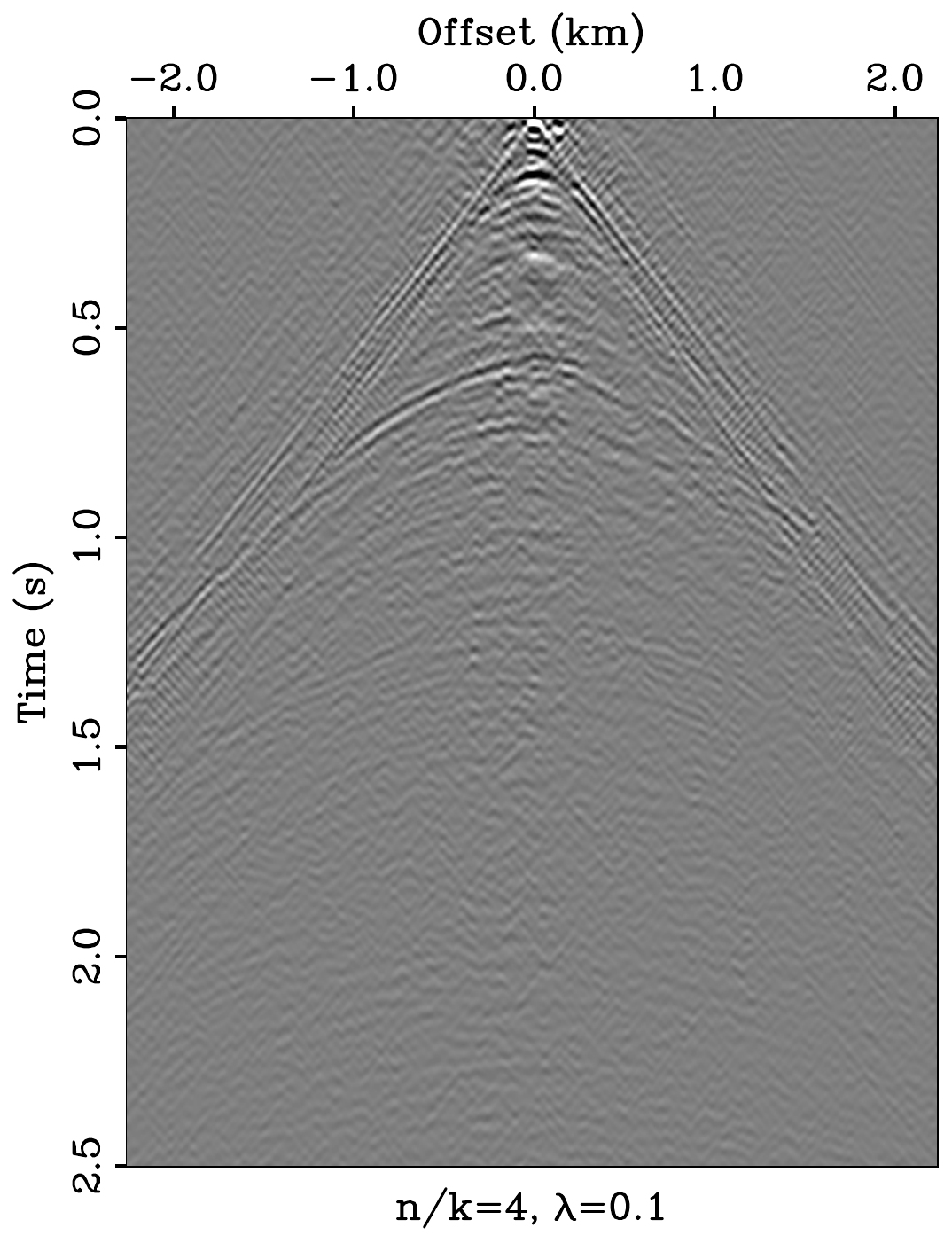}
      \includegraphics[width=0.4\textwidth]{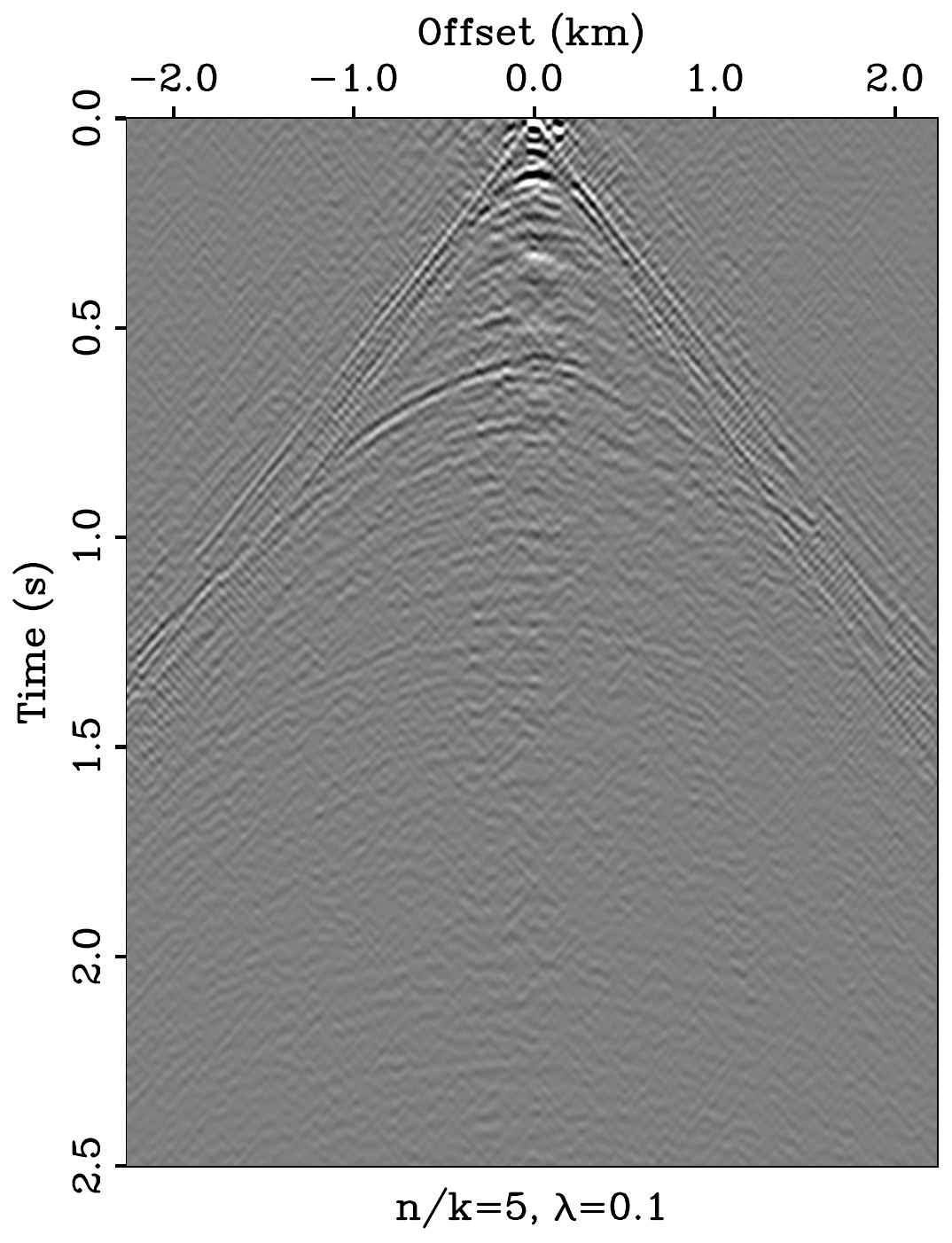}
      \includegraphics[width=0.4\textwidth]{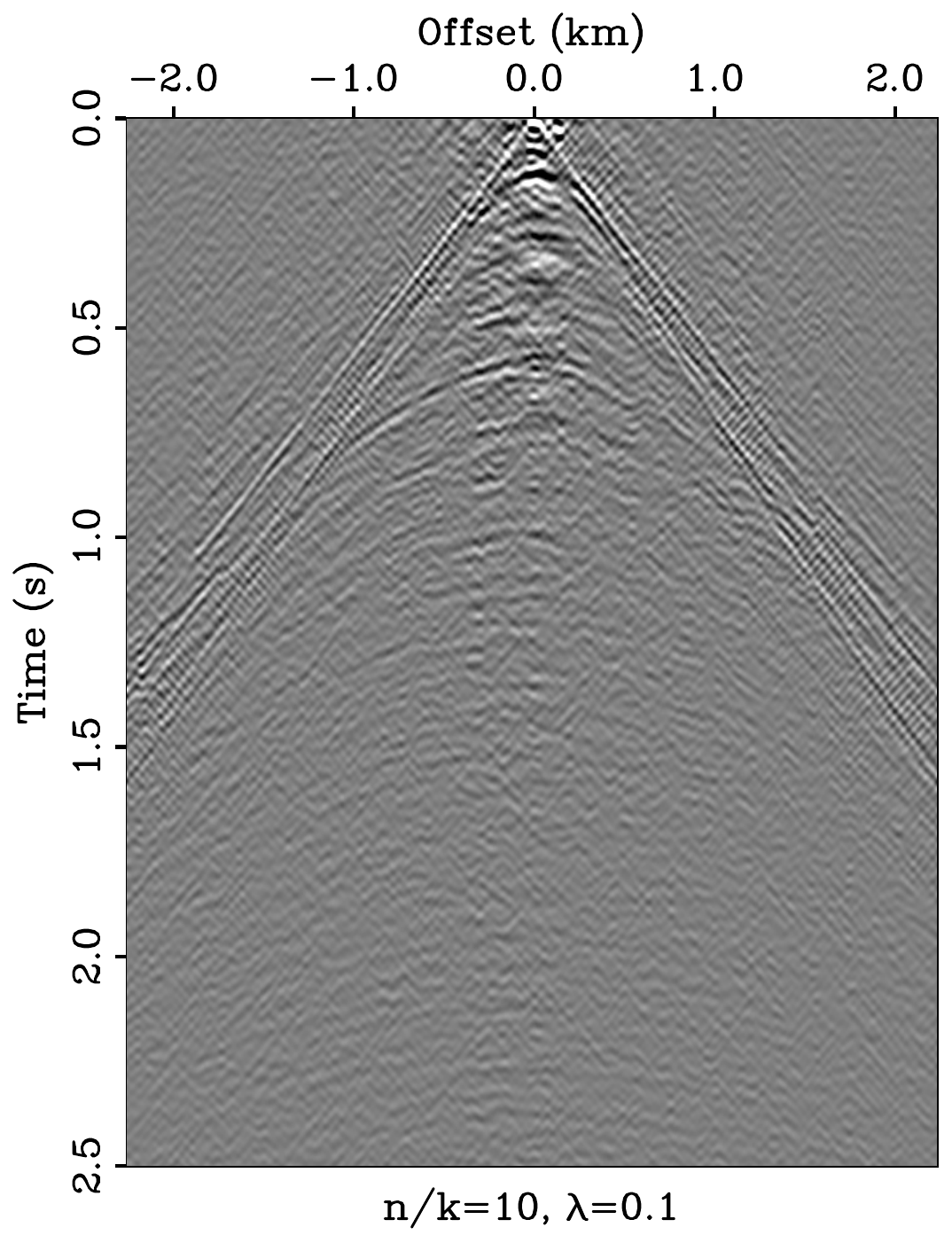}
    \caption{Results of MDD with the proposed low-rank regularization.}
    \label{fig:Volve_QQtMDD}
\end{figure}

Figure \ref{fig:Volve_QQtMDD} showcases the time-offset domain results from MDD using our proposed reciprocal low-rank factorization with varying rank values: $k=60$, $k=45$, $k=30$, and $k=18$. In all instances, we maintain the same regularization parameter of $\lambda=0.1$. The results with $k=60$, $k=45$, and $k=30$ are comparable with the symmetric reference. However, they allow a reduction in the memory footprint of the solution by a factor of $n_r/k=3$, $n_r/k=4$, and $n_r/k=5$, respectively. The result with $k=18$ achieves acceptable results, where the key events are still successfully retrieved, however, some additional noise starts to appear in the solution. On the other hand, the memory footprint is reduced by a factor of $n_r/k=10$ compared to the scheme to generate the benchmark solution.

\section*{Discussion}
Multi-dimensional deconvolution has been for a long time an elusive dream of many geophysicists; this is particularly the case for large-scale 3D applications, mostly due to the extreme computational cost and memory footprint of this algorithm, among other practical issues related to the sampling of the acquisition geometry. Whilst recent publications have reported the successful application of MDD to 3D datasets \citep{Boieroetal2023, Haacke2023, Ltaief2023}, the quest for a faster and cheaper solution is still relevant. The new parametrization of the sought-after Green's function proposed in this work may reveal the key to improving the scalability of MDD to even larger datasets. Despite being so far only focused on 2D scenarios, our work has shown that we can store the entire information of the matrix $\mathbf{X}$ into a single factor $\mathbf{Q}$ that is significantly smaller in size (up to 5 in the field data example) with minimal degradation on the quality of the solution. As a by-product, this parametrization also allows for a faster application of the forward and adjoint modelling operators. 

In 3D scenarios, however, we expect the solution not to be globally low-rank, similar to what was previously observed for the kernel of the modelling operator $D$ \citep{hong2021, Yuxi2022IJHPCA, Ravasietal2022EAGE}. As such, our proposed method is required to accommodate for an unknown Green's function that assumes a blocked low-rank structure. Extending our proposed method to address this scenario is a challenging and ongoing area of investigation. Furthermore, it holds potential for integration with compressed operators, particularly by utilizing a tile low-rank strategy \cite[]{Ravasietal2022EAGE,Ltaief2023}

Finally, another promising research avenue involves the exploration of an adaptive scheme for determining the optimal value of $k$ at different frequencies, considering the increased rank deficiency typically observed at lower frequencies.

\section*{Conclusion}
We have presented a new factorization-based, low-rank regularization for MDD that inherently fulfills the reciprocity property of Green's functions retrieved along the target datum. We employed the accelerated proximal gradient method to solve MDD with the proposed parametrization of the sought-after wavefield in which the Frobenius norm is further introduced to constrain the magnitude of the low-rank matrix.

Demonstrations using synthetic and field data illustrate that our approach enhances frequency-domain MDD in two key ways: (1) it yields results comparable to those of the multi-frequency method while significantly reducing memory requirements, making it a promising solution for large-scale 3D MDD applications (2) it demonstrates robustness on par with the preconditioned multi-frequency method when the up- and down-going wavefield contains spurious artifacts.

\section*{Acknowledgments}
For computer time, this research leveraged the resources of the Supercomputing Laboratory at King Abdullah University of Science \& Technology (KAUST) in Thuwal, Saudi Arabia. We sincerely thank Ning Wang (KAUST) for her assistance in preparing the data for our first example, and David Vargas (Utrecht University) for providing the salt dataset. We also acknowledge the use of ChatGPT in the process of writing the paper.

\appendix
\section*{The gradient of misfit with respect to the low-rank factors}
\label{gradLandR}
Here we present detailed derivations of the partial derivative of $\displaystyle f_2(\mathbf{L},\mathbf{R})$ with $\mathbf{L}$ and $\mathbf{R}$, mostly based on the property that the trace of a matrix is invariant under circular shifts.
First, let us rewrite $f_2(\mathbf{L},\mathbf{R})=(1/2)\mathrm{tr}\big((\mathbf{M}-\mathbf{U})^\mathrm{H}(\mathbf{M}-\mathbf{U})\big)$, where $\mathbf{M}=\mathbf{D}\mathbf{L}\mathbf{R}$. Then we have 
\begin{subequations}
\renewcommand{\theequation}{\theparentequation.\arabic{equation}}
\begin{align}
f_2(\mathbf{L},\mathbf{R})&=(1/2)\mathrm{tr}\big(\mathbf{M}^\mathrm{H} \mathbf{M}\big)-\mathrm{tr}\big(\mathbf{M}^\mathrm{H} \mathbf{U}\big), \\ 
&=(1/2)\mathrm{tr}\big(\mathbf{R}^\mathrm{H} \mathbf{L}^\mathrm{H}\mathbf{D}^\mathrm{H} \mathbf{D}\mathbf{L}\mathbf{R}\big)- \mathrm{tr}\big(\mathbf{R}^\mathrm{H} \mathbf{L}^\mathrm{H}\mathbf{D}^\mathrm{H}\mathbf{U}\big),
\end{align}
\end{subequations}
where we have dropped one term, which is independent on both $\mathbf{L}$ and $\mathbf{R}$. Following the cyclic property of matrices, we can write 
\begin{subequations}
\label{eq:shifts_1}
\renewcommand{\theequation}{\theparentequation.\arabic{equation}}
\begin{align}
    \mathrm{tr}\big(\mathbf{R}^\mathrm{H} \mathbf{L}^\mathrm{H}\mathbf{D}^\mathrm{H} \mathbf{D}\mathbf{L}\mathbf{R}\big)=    \mathrm{tr}\big(\mathbf{L}^\mathrm{H}\mathbf{D}^\mathrm{H} \mathbf{D}\mathbf{L}\mathbf{R}\mathbf{R}^\mathrm{H}\big)\\
    \mathrm{tr}\big(\mathbf{R}^\mathrm{H} \mathbf{L}^\mathrm{H}\mathbf{D}^\mathrm{H} \mathbf{D}\mathbf{L}\mathbf{R}\big)=     \mathrm{tr}\big(\mathbf{L}\mathbf{R}\mathbf{R}^\mathrm{H} \mathbf{L}^\mathrm{H}\mathbf{D}^\mathrm{H} \mathbf{D}\big)
\end{align}
\end{subequations}
The motivation for applying such shifts in equation \ref{eq:shifts_1} is to be able to evaluate the contribution of a change in $\mathrm{tr}\big(\mathbf{R}^\mathrm{H}\mathbf{L}^\mathrm{H}\mathbf{D}^\mathrm{H} \mathbf{D}\mathbf{L}\mathbf{R}\big)$ associated with a change in $\mathbf{L}$. Following the identity \begin{subequations}
    \begin{align}
        \frac{\partial\mathrm{tr}(\mathbf{X}^\mathrm{H}\mathbf{Y})}{\partial\mathbf{X}}&=\mathbf{Y},\\
        \frac{\partial \mathrm{tr}(\mathbf{X}\mathbf{Y})}{\partial\mathbf{X}}&=\mathbf{Y}^\mathrm{H},
    \end{align}
\end{subequations}
where we assume that the dimensions of matrices are compatible, we have
\begin{subequations}
\renewcommand{\theequation}{\theparentequation.\arabic{equation}}
    \begin{align}
\frac{\partial \mathrm{tr}\big(\mathbf{L}^\mathrm{H}\mathbf{D}^\mathrm{H} \mathbf{D}\mathbf{L}\mathbf{R}\mathbf{R}^\mathrm{H}\big)}{\partial\mathbf{L}} &=\mathbf{D}^\mathrm{H} \mathbf{D}\mathbf{L}\mathbf{R}\mathbf{R}^\mathrm{H}, \\
\frac{\partial \mathrm{tr}\big(\mathbf{L}\mathbf{R}\mathbf{R}^\mathrm{H} \mathbf{L}^\mathrm{H}\mathbf{D}^\mathrm{H} \mathbf{D})}{\partial\mathbf{L}} &=\big(\mathbf{R}\mathbf{R}^\mathrm{H} \mathbf{L}^\mathrm{H}\mathbf{D}^\mathrm{H} \mathbf{D}\big)^\mathrm{H}. 
\end{align}
\end{subequations}
Finally, we obtain 
\begin{subequations}
\begin{align}
    \frac{\partial \mathrm{tr}\big(\mathbf{R}^\mathrm{H} \mathbf{L}^\mathrm{H}\mathbf{D}^\mathrm{H} \mathbf{D}\mathbf{L}\mathbf{R}\big)}{\partial\mathbf{L}} &= \mathbf{D}^\mathrm{H} \mathbf{D}\mathbf{L}\mathbf{R}\mathbf{R}^\mathrm{H}+\big(\mathbf{R}\mathbf{R}^\mathrm{H} \mathbf{L}^\mathrm{H}\mathbf{D}^\mathrm{H} \mathbf{D}\big)^\mathrm{H},\\
    &=2\mathbf{D}^\mathrm{H} \mathbf{D}\mathbf{L}\mathbf{R}\mathbf{R}^\mathrm{H}.
\end{align}
\end{subequations}
Following the same idea, we can have
\begin{equation}
    \frac{\partial     \mathrm{tr}\big(\mathbf{R}^\mathrm{H} \mathbf{L}^\mathrm{H}\mathbf{D}^\mathrm{H}\mathbf{U}\big)}{\partial \mathbf{L}}=\frac{\partial     \mathrm{tr}\big(\mathbf{L}^\mathrm{H}\mathbf{D}^\mathrm{H}\mathbf{U}\mathbf{R}^\mathrm{H} \big)}{\partial \mathbf{L}}=\mathbf{D}^\mathrm{H}\mathbf{U}\mathbf{R}^\mathrm{H}.
\end{equation}
Therefore,
\begin{subequations}
\begin{align}
    \frac{\partial f_2}{\partial \mathbf{L}}&=\mathbf{D}^\mathrm{H}\mathbf{D}\mathbf{L}\mathbf{R}\mathbf{R}^\mathrm{H}-\mathbf{D}^\mathrm{H}\mathbf{U}\mathbf{R}^\mathrm{H}, \\
    &=\mathbf{D}^\mathrm{H}(\mathbf{D}\mathbf{L}\mathbf{R}-\mathbf{U})\mathbf{R}^\mathrm{H}.
\end{align}
\end{subequations}
We can also follow a similar derivation to obtain the derivative of $f(\mathbf{L},\mathbf{R})$ with $\mathbf{R}$ as below:
\begin{subequations}
\renewcommand{\theequation}{\theparentequation.\arabic{equation}}
\begin{align}
f_2(\mathbf{L},\mathbf{R})&=(1/2)\mathrm{tr}\big((\mathbf{D}\mathbf{L}\mathbf{R})^\mathrm{H} (\mathbf{D}\mathbf{L}\mathbf{R})\big)-\mathrm{tr}\big((\mathbf{D}\mathbf{L}\mathbf{R})^\mathrm{H} \mathbf{U}\big), \\ 
&=(1/2)\mathrm{tr}\big(\mathbf{R}^\mathrm{H} (\mathbf{D}\mathbf{L})^\mathrm{H} (\mathbf{D}\mathbf{L})\mathbf{R}\big)-\mathrm{tr}\big(\mathbf{R}^\mathrm{H} (\mathbf{D}\mathbf{L})^\mathrm{H}\mathbf{U}\big).
\end{align}
\end{subequations}
Based on the identity 
\begin{equation}
    \frac{\partial \mathrm{tr}\big(\mathbf{X}^\mathrm{H}\mathbf{D}\mathbf{X}\big)}{\partial \mathbf{X}}=\mathbf{DX}+\mathbf{D}^\mathrm{H}\mathbf{X},
\end{equation}
we obtain 
\begin{equation}
    \frac{\partial \mathrm{tr}\big(\mathbf{R}^\mathrm{H} (\mathbf{D}\mathbf{L})^\mathrm{H} (\mathbf{D}\mathbf{L})\mathbf{R}\big)}{\partial \mathbf{R}}=2\mathbf{(\mathbf{D}\mathbf{L})^\mathrm{H} (\mathbf{D}\mathbf{L})}\mathbf{R}.
\end{equation}
Therefore,
\begin{subequations}
\begin{align}
    \frac{\partial f_2}{\partial \mathbf{R}}&=\mathbf{(\mathbf{D}\mathbf{L})^\mathrm{H} (\mathbf{D}\mathbf{L})}\mathbf{R}-(\mathbf{D}\mathbf{L})^\mathrm{H}\mathbf{U}, \\
    &=\mathbf{(\mathbf{D}\mathbf{L})^\mathrm{H} (\mathbf{D}\mathbf{L}}\mathbf{R}-\mathbf{U}).
\end{align}
\end{subequations}

\bibliographystyle{seg}  
\bibliography{references}

\end{document}